\documentclass[12pt]{article}
\usepackage{graphicx,cite}
\usepackage{fullpage}
\usepackage{pmat}
\usepackage{amsmath,amsthm,amsfonts,amssymb}
\usepackage{graphicx}
\usepackage{color}
\usepackage[all]{xy}  

\newcommand{\NC}{noncommutative}

\newcommand{\ba}{\begin{array}}
\newcommand{\ea}{\end{array}}

\newcommand{\nn}{\nonumber\\}

\newcommand{\C}{{\bf C}}

\newcommand{\del}{\partial}

\newcommand{\scr}{\scriptsize}

\newcommand{\at}{\tilde{a}}
\newcommand{\bt}{\tilde{b}}
\newcommand{\ct}{\tilde{c}}
\newcommand{\dt}{\tilde{d}}

\newcommand{\zt}{\tilde{z}}
\newcommand{\wt}{\tilde{w}}

\makeatletter
\def\section{\@startsection {section}{1}{\z@}{-1.5ex plus-1ex minus
   -.2ex}{1.0ex plus.2ex}{\reset@font\large\bf}}
\def\subsection{\@startsection{subsection}{2}{\z@}{-3.25ex plus-1ex
    minus-.2ex}{0.75ex plus.2ex}{\reset@font\it}}

\def\@sect#1#2#3#4#5#6[#7]#8{
\ifnum #2>\c@secnumdepth
    \let\@svsec\@empty\else
    \refstepcounter{#1}{\centering%Added centering group from
    \edef\@svsec{\csname the#1\endcsname\ifnum #2=1
    .\else\relax\fi\hskip .5em}\fi
    \@tempskipa #5\relax
     \ifdim \@tempskipa>\z@
       \begingroup #6\relax
         \@hangfrom{\hskip #3\relax\@svsec}{\interlinepenalty \@M
    #8\par}%
       \endgroup%
       }%...to here
      \csname #1mark\endcsname{#7}\addcontentsline
        {toc}{#1}{\ifnum #2>\c@secnumdepth \else
                     \protect\numberline{\csname the#1\endcsname}\fi
                   #7}\else
       \def\@svsechd{#6\hskip #3\relax  %% \relax added 2 May 90
                  \@svsec #8\csname #1mark\endcsname
                     {#7}\addcontentsline
                          {toc}{#1}{\ifnum #2>\c@secnumdepth \else
                            \protect\numberline{\csname
    the#1\endcsname}\fi
                      #7}}\fi
    \@xsect{#5}}
\def\@cite#1#2{({#1\if@tempswa ; #2\fi})}
\def\@biblabel#1{\relax}
\def\@citex[#1]#2{\if@filesw\immediate\write\@auxout{\string\citation{#2}}\fi
 \let\@citea\@empty
 \@cite{\@for\@citeb:=#2\do
   {\@citea\def\@citea{,\penalty\@m\ }%
    \def\@tempa##1##2\@nil{\edef\@citeb{\if##1\space##2\else##1##2\fi}}%
    \expandafter\@tempa\@citeb\@nil
    \@ifundefined{b@\@citeb}{{\reset@font\bf ?}\@warning
      {Citation `\@citeb' on page \thepage \space undefined}}%
    {\csname b@\@citeb\endcsname}}}{#1}}
\long\def\@makecaption#1#2{%
 \vskip\abovecaptionskip
 \sbox\@tempboxa{#1. #2}%
 \ifdim \wd\@tempboxa >\hsize
   #1: #2\par
 \else
   \global \@minipagefalse
   \hb@xt@\hsize{\hfil\box\@tempboxa\hfil}%
 \fi
 \vskip\belowcaptionskip}
\makeatother

\begin{document}

\begin{titlepage}
\null
\begin{flushright}
Proc.Roy.Soc.A{\bf 465} (2009) 2613% -- 2632
\\
arXiv:0812.1222
\\
December, 2008
\end{flushright}
\vskip 1cm
\begin{center}

 \vskip 1.5cm

{\large \bf B\"acklund Transformations and the Atiyah-Ward Ansatz for}

 \vskip 0.3cm

{\large \bf Noncommutative Anti-Self-Dual Yang-Mills Equations}

\vskip 1.5cm

{\large Claire R. Gilson$^\dagger$, 
 %\footnote{E-mail: }
Masashi Hamanaka$^{\ddagger}$
and
Jonathan J. C. Nimmo$^\dagger$
 %\footnote{E-mail: }
 }

\vskip 7mm

        {\it $^\dagger$Department of Mathematics, University of
 Glasgow\\
              Glasgow G12 8QW, UK}

\vskip 0.5cm

        {\it $^\ddagger$Department of Mathematics, University of
 Nagoya,\\
                      Nagoya, 464-8602, JAPAN}

%\date

\vskip 1.5cm

{\bf Abstract}
\end{center}
\baselineskip 6mm
{%\small
We present B\"acklund transformations for
the noncommutative anti-self-dual Yang-Mills equation 
where the gauge group is $G=GL(2)$ and use it to generate a
series of exact solutions from a simple seed solution. 
The solutions generated by this approach are 
represented in terms of quasideterminants.
We also explain the origins of all of the ingredients of the B\"acklund
transformations within the framework of noncommutative twistor theory. 
In particular we show that the generated solutions 
belong to a noncommutative version of the Atiyah-Ward ansatz.
%In the commutative limit, our results coincide with
%those by Corrigan, Fairlie, Yates and Goddard.
}
%\vskip 0.1cm

\begin{center}
{\bf Key Words}
\end{center}
noncommutative integrable systems, anti-self-dual Yang-Mills equation,
quasideterminant solutions, Atiyah-Ward ansatz, Penrose-Ward
 transformation

\end{titlepage}
\clearpage
\baselineskip 6mm

\section{Introduction}

In both mathematics and physics, a noncommutative extension is a natural generalization of a commutative theory that sometimes leads to a new and deeper understanding of that theory.
While matrix (or more general, non-Abelian) generalizations have been studied 
for a long time, the generalization to \NC{} flat spaces,
triggered by developments in string theory (see e.g. Douglas and
Nekrasov (2002); Szabo (2003)), has become a hot topic recently. These generalizations are realized by replacing all products in the commutative theory with associative but noncommutative Moyal products. In gauge theories, such a \NC{} extension is equivalent to the
presence of a background magnetic field. Many successful applications of the analysis of \NC{} solitons to $D$-brane dynamics have been made. 
(Here we use the word ``soliton'' as a stable configuration 
which possesses localized energy densities and hence includes static
configurations such as instantons.)

Integrable systems and soliton theory can also be extended 
to a \NC{} setting and yield interesting results and applications. 
(For reviews, see e.g. 
Dimakis and M\"uller-Hoissen (2004); Hamanaka (2003); Hamanaka (2005a);
Hamanaka and Toda (2003); Kupershmidt (2000); Lechtenfeld (2008); 
Mazzanti (2007); Tamassia (2005).)
Among them, the \NC{} anti-self-dual Yang-Mills (ASDYM) equation in 4-dimensions
is important because in the Euclidean signature $(+$$+$$+$$+)$,  the
ADHM construction can be used to find all exact instanton solutions and gives rise
to new physical objects such as $U(1)$ instantons (Nekrasov and Schwarz (1998)).
In the split signature $(+$$+$$-$$-)$, many \NC{} integrable equations can be 
derived from the \NC{} ASDYM equation by a reduction process
(See Hamanaka (2005b); Hamanaka (2006) and references therein).
%Since the ASDYM equation is in a gauge theory, 
%the Moyal deformation of it has a physical correspondence. Furthermore, 
Integrable aspects of the \NC{} ASDYM equation
can be understood in the geometrical framework of \NC{} twistor theory
(Brain (2005); Brain and Majid (2008); Hannabuss (2001); 
Horv\'ath, Lechtenfeld and Wolf (2002); Ihl and Uhlmann (2003);
Kapustin, Kuznetsov and Orlov (2001); Lechtenfeld and  Popov (2002); Takasaki (2001)).
Therefore it is worth studying the integrable aspects of the
\NC{} ASDYM equation in detail both for the applications 
to lower-dimensional integrable equations 
and to the corresponding physical situations 
in the framework of a \NC{} analogue of $N=2$ string theory
(Lechtenfeld, Popov and Spendig (2001a); Lechtenfeld, Popov and Spendig
(2001b)). Here solitons are not, in general, static and 
suggest the existence of some kinds of new configurations.
For these purposes, 
B\"acklund transformations play an important role
in constructing exact solutions and revealing an
(infinite-dimensional) symmetry of the solution space 
in terms of the transformation group. Also, a 
twistor description is useful for a discussion of the  
origin of the transformations and for checking
whether or not the group action is transitive.
%Hence it is worth studying B\"acklund transformations
%for \NC{} ASDYM equation for application of
%exact solutions to physics and discussion on integrablility of it.

In the present paper, we give B\"acklund transformations for the
\NC{} ASDYM equation where the gauge group is $G=GL(2)$ and use them to generate a
sequence of exact solutions from a simple seed solution. 
This approach gives both finite action solutions (instantons) and infinite action solutions (such as nonlinear plane waves). The solutions obtained are written
in terms of quasideterminants (Gelfand and Retakh (1991); 
Gelfand and Retakh (1992)) which
appear also in the construction 
of exact soliton solutions in lower-dimensional \NC{} integrable equations 
such as 
the Toda equation
(Etingof, Gelfand and Retakh (1997);
Etingof, Gelfand and Retakh (1998);
Li and Nimmo (2008);
Li and Nimmo (2009)),
the KP and KdV equations 
(Dimakis and M\"uller-Hoissen (2007);
Etingof, Gelfand and Retakh (1997);
Gilson and Nimmo (2007);
Hamanaka (2007)), 
the Hirota-Miwa equation 
(Gilson, Nimmo and Ohta (2007); 
Li, Nimmo and Tamizhmani (2009); Nimmo (2006)),
the mKP equation (Gilson, Nimmo and Sooman (2008a); 
Gilson, Nimmo and Sooman (2008b)),
the Schr\"odinger equation
(Goncharenko and Veselov (1998);
Samsonov and Pecheritsin (2004)),
the Davey-Stewartson equation
(Gilson and Macfarlane (2009)),
the dispersionless equation
(Hassan (2009)),
and the chiral model
(Haider and Hassan (2008)),
where they play the role that
determinants do in the corresponding commutative integrable systems.
We also clarify the origin of the results from the viewpoint of
\NC{} twistor theory by using \NC{} Penrose-Ward correspondence or
by solving a \NC{} Riemann-Hilbert problem.
It is shown that the solutions generated 
belong to a noncommutative version 
of the {\it Atiyah-Ward ansatz} (Atiyah and Ward (1977)).
%This is a extended version of a previous paper 
%(Gilson, Hamanaka and Nimmo (2008)), 
%which does not describe the connections to twistor theory. 
%The present paper contains results from the earlier one 
%in order that is more self-contained.

The discussion and strategy used in this paper are simple \NC{} generalizations 
of those used in the commutative case (Corrigan, Fairlie, Yates and Goddard (1978);
Mason, Chakravarty and Newman 
(1988); Mason and Woodhouse (1996)). 
In the commutative limit, our results coincide in part 
with the known results but in the noncommutative case,  
there are several nontrivial points.
Firstly, in Section 3.2 we show that
quasideterminants are ideally suited to the \NC{} extension of the known results
and greatly simplify the proofs of the B\"acklund transformations 
even in the commutative limit. 
The simple quasideterminant representations of Yang's $J$-matrix
are new and imply the important result that the B\"acklund transformation is not
just a gauge transformation.
It is possible for the \NC{} twistor description to work as it does in the commutative setting 
because one of the three local coordinates can be taken to be a commutative variable. 
%We note that we have to introduce one additional
%degree of freedom as in Eq. (\ref{param})
%because in \NC{} gauge theory, $U(1)$ part of gauge groups 
%is crucial. This point actually corresponds to the existence of
%new physical objects such as $U(1)$ instantons.]  

In our treatment, all dependent variables belong to a ring, 
which has an associative but not necessarily commutative product.
%that is, 
%all multiplication is associative but is not assumed to be commutative. 
Hence the results we obtain are available %not only 
in any \NC{} settings such as the Moyal-deformed, 
%for discussion on \NC{} flat spaces 
%which are realized by replacement of all products with Moyal-products, 
%but also in other \NC{} settings such as 
matrix or quaternion-valued ASDYM equations.

\section{The \NC{} ASDYM equation}

Let us consider \NC{} Yang-Mills theories in 
$4$-dimensions where the gauge group is $GL(N)$
and the real coordinates are $x^\mu,~\mu=0,1,2,3$.
In the rest of the paper, we follow the conventions 
of notation given in Mason and Woodhouse (1996).

\subsection{The \NC{} ASDYM equation}

The \NC{} ASDYM equation is derived from the compatibility condition
of the linear system
\begin{eqnarray}
L\psi&:=&(D_w-\zeta D_{\tilde{z}})\psi=
 \left(\del_w+A_w-\zeta (\del_{\zt}+A_{\tilde{z}})\right)\psi%(x;\zeta)
= 0,\nn
M\psi&:=&(D_z-\zeta D_{\tilde{w}})\psi=
  \left(\del_z+A_z-\zeta (\del_{\wt}+A_{\tilde{w}})\right)\psi%(x;\zeta)
= 0,
\label{lin_asdym}
\end{eqnarray}
where $A_z,A_w,A_{\tilde{z}},A_{\tilde{w}}$
and $D_z,D_w,D_{\tilde{z}},D_{\tilde{w}}$
denote gauge fields and covariant derivatives in Yang-Mills theory,
respectively. The (commutative) variable $\zeta$ is a local coordinate
of a one-dimensional complex projective space $\mathbb{C}P_1$,
and is called the {\it spectral parameter}. We note that $\psi$
is not regular at $\zeta=\infty$ because if it were
regular, by Liouville's theorem it would be a constant fuction
and the gauge fields would be flat (see e.g. Mason and Woodhouse (1996)).

The four complex coordinates $z,\zt,w,\wt$
are double null coordinates (Mason and Woodhouse (1996)).
By imposing the corresponding reality conditions,
we can realize real spaces with different signatures, that is,
\begin{itemize}
 \item the Euclidean space, obtained by putting $\bar{w}=-\wt; \bar{z}=\zt$,
for example,
       \begin{eqnarray}
        \left[\begin{array}{cc}\tilde{z}&w\\\tilde{w}&z\end{array}\right]
        =\frac{1}{\sqrt{2}}
         \left[\begin{array}{cc}x^0+ix^1&-(x^2-ix^3)\\x^2+ix^3&x^0-ix^1
               \end{array}\right],
       \end{eqnarray}

\item the Ultrahyperbolic space, obtained by putting $\bar{w}=\wt; \bar{z}=\zt$,
for example,
       \begin{eqnarray}
        \left[\begin{array}{cc}\tilde{z}&w\\\tilde{w}&z\end{array}\right]
        =\frac{1}{\sqrt{2}}
         \left[\begin{array}{cc}x^0+ix^1&x^2-ix^3\\x^2+ix^3&x^0-ix^1
               \end{array}\right]
~~~\mbox{or}~~~z,w,\tilde{z},\tilde{w}\in \mathbb{R}.
       \end{eqnarray}
\end{itemize}

The compatibility condition $[L,M]=0$, gives rise to
a quadratic polynomial in $\zeta$ and each coefficient
yields the \NC{} ASDYM equations, with explicit representations
\begin{eqnarray}
 F_{wz}&=&\del_{w} A_z -\del_{z} A_w+[A_w,A_z]=0,\nn
 F_{\wt\zt}&=&\del_{\tilde{w}} A_{\tilde{z}} -\del_{\tilde{z}} A_{\tilde{w}}
 +[A_{\tilde{w}},A_{\tilde{z}}]=0,\nn
 F_{z\zt}-F_{w\wt}&=&\del_{z} A_{\tilde{z}} -\del_{\tilde{z}} A_{z}
 +\del_{\tilde{w}} A_{w} -\del_{w} A_{\tilde{w}}
 +[A_z,A_{\tilde{z}}]
 -[A_{w},A_{\tilde{w}}]=0,
\label{asdym}
\end{eqnarray}
which is equivalent to the ASD condition for gauge fields $F_{\mu\nu}=-*F_{\mu\nu}$ in the real representation where the symbol $*$ is the Hodge dual operator.

When the compatibility conditions are satisfied, the linear system \eqref{lin_asdym} has $N$ independent solutions. Hence the solution $\psi(x;\zeta)$ can be interpreted as an $N\times N$ matrix whose columns are the $N$ independent solutions.

Gauge transformations act on the linear system (\ref{lin_asdym}) as
\begin{eqnarray}
 L\mapsto g^{-1}Lg,~
 M\mapsto g^{-1}Mg,~
 \psi\mapsto g^{-1}\psi,~~~g\in G.
\end{eqnarray}

\subsection{The \NC{} Yang equation and $J,K$-matrices}

Here we discuss the different potential forms of the \NC{} ASDYM equations
such as the \NC{} $J,K$-matrix formalisms and the \NC{} Yang equation,
which were already presented by e.g. Takasaki (2001).

Let us first discuss the {\it $J$-matrix formalism}
of the \NC{} ASDYM equation (\ref{asdym}).
The first and second equations of (\ref{asdym}) 
are the compatibility conditions of 
\begin{eqnarray}
\label{h}
D_zh=0,~D_wh=0,~~~
\mbox{and}~~~ 
\label{ht}
D_{\zt}\tilde{h}=0,~D_{\wt} \tilde{h}=0, 
\end{eqnarray}
respectively.
Here $h$ and $\tilde{h}$ are $N\times N$ matrices, 
whose $N$ columns of $h$ and $\tilde{h}$ are 
independent solutions of the linear systems.
The existence of them is formally proved 
in the case of the Moyal deformation by Takasaki (2001)
and presumed here.
These equations can be satisfied by choosing 
\begin{eqnarray}
\label{a}
A_{z}=-(\del_z h)h^{-1}, ~~~  A_{w}=-(\del_w h)h^{-1},~~~
\label{at}
A_{\zt}=-(\del_{\zt}\tilde{h})\tilde{h}^{-1}, ~~~
A_{\wt}=-(\del_{\wt}\tilde{h})\tilde{h}^{-1}.
\end{eqnarray}

By defining $J=\tilde{h}^{-1}h $, 
the third equation of (\ref{asdym}) becomes
\begin{eqnarray}
\label{yang}
 \del_z(J^{-1} \del_{\zt} J)-\del_w (J^{-1}\del_{\wt} J)=0.
\end{eqnarray}
%or equivalently,
%\begin{eqnarray}
% \del\left[J^{-1}\delt J\right]\wedge \omega=0.
%\end{eqnarray}
%where $\del=dw \del_w +dz\del_z,~\delt=d\wt \del_{\wt} +d\zt\del_{\zt}$
%$\omega$ is the same one in Eq. (\ref{omega}). 
This equation is called the {\it \NC{} Yang equation} 
and the matrix $J$ is called {\it Yang's $J$-matrix}.
 
Gauge transformations act on $h$ and $\tilde{h}$ as
\begin{eqnarray}
 h\mapsto g^{-1}h,~
 \tilde{h}\mapsto g^{-1}\tilde{h},~~~g\in G.
\end{eqnarray}
Hence Yang's $J$-matrix is gauge invariant.
Gauge fields are obtained from a solution $J$ of the \NC{} Yang's
equation via a decomposition $J=\tilde{h}^{-1} h$, and \eqref{a}. 
The different decompositions correspond to different choices of gauge.

%\begin{eqnarray*}
%A_{z}=-(\partial_z  h) h^{-1}, ~~~  A_{w}=-(\partial_w h) h^{-1},~~~
%A_{\tilde{z}}=-(\partial_{\tilde{z}}\tilde{h}) \tilde{h}^{-1}, ~~~
%A_{\tilde{w}}=-(\partial_{\tilde{w}}\tilde{h}) \tilde{h}^{-1}.
%\end{eqnarray*} 

There is another potential form of the \NC{} ASDYM equation,
known as the {\it $K$-matrix formalism}. In the gauge in which $A_w=A_z=0$,
the third equation of (\ref{asdym})
becomes $\del_z A_{\zt}-\del_w A_{\wt}=0$.
This implies the existence of a potential $K$ such
that $A_{\zt}=\del_{w}K,A_{\wt}=\del_{z}K$.
Then the second equation of (\ref{asdym}) becomes
\begin{eqnarray}
 \del_z\del_{\zt}K -\del_w\del_{\wt}K +[\del_w K, \del_z K]=0.
\end{eqnarray}
This gauge is suitable for the discussion of 
(binary) Darboux transformations for (\NC{}) ASDYM equation 
(Gilson, Nimmo and Ohta (1998); Nimmo, Gilson and Ohta (2000);
Saleem, Hassan and Siddiq (2007)).

\section{B\"acklund transformation for the \NC{} ASDYM equation}

In this section, we present two kind of B\"acklund transformations
which leave the \NC{} Yang equation for $G=GL(2)$ invariant.
This is a \NC{} version of the Corrigan-Fairlie-Yates-Goddard  
transformation (Corrigan, Fairlie, Yates and Goddard (1978)).
This transformation generates a class of exact solutions 
which belong to a \NC{} version 
of the {\it Atiyah-Ward ansatz} (Atiyah and Ward (1978)) labeled by
a nonnegative integer $l\in \mathbb{Z}_{\geq 0}$.
The origin of these results will be clarified in the next section.
%This would relates to \NC{} B\"acklund transformations 
%for \NC{} integrable equations in lower dimensions via reductions.

In order to discuss B\"acklund transformations for
the \NC{} Yang equation, we parameterize
the $2\times 2$ matrix $J$ as
\begin{eqnarray}
\label{param}
 J=\left[\begin{array}{cc} f -g b^{-1} e&-g b^{-1}
   \\ b^{-1}  e &b^{-1}\end{array}
   \right].
\end{eqnarray}
This parameterization is always possible
when $f$ and $b$ are invertible. 
In contrast with the commutative case, where only $f$ appears, 
in the \NC{} setting, we need to introduce another variable $b$.
In the commutative limit we may choose $b=f$. 

Then the \NC{} Yang equation (\ref{yang}) is decomposed as
\begin{eqnarray}
  \partial_{z}(f^{-1} g_{\tilde{z}}  b^{-1})-\partial_{w}(f^{-1} g_{\tilde{w}}
  b^{-1})=0,~~~
  \partial_{\tilde{z}}(b^{-1} e_z  f^{-1})
  -\partial_{\tilde{w}}(b^{-1} e_w f^{-1})&=&0,\nonumber\\
  \partial_{z}( b_{\tilde{z}} b^{-1})-
  \partial_w(b_{\tilde{w}} b^{-1})
  -e_z f^{-1} g_{\tilde{z}} b^{-1}
  +e_w f^{-1} g_{\tilde{w}} b^{-1}&=&0,\nonumber\\
  \partial_{z}(f^{-1} f_{\tilde{z}})-
  \partial_{w}(f^{-1} f_{\tilde{w}})
  -f^{-1} g_{\tilde{z}} b^{-1} e_{z}
  +f^{-1} g_{\tilde{w}} b^{-1} e_{w}&=&0,
\label{dYang}
\end{eqnarray}
where subscripts denote partial derivatives.

\subsection{The \NC{} Corrigan-Fairlie-Yates-Goddard transformation}

The \NC{} Corrigan-Fairlie-Yates-Goddard transformation is
a composition of the following two B\"acklund transformations
for the \NC{} Yang equations (\ref{dYang}).

\begin{itemize}
\item $\beta$-transformation (Mason and Woodhouse (1996)): 
\begin{eqnarray}
 \label{new}
   e_w^{\mbox{\scriptsize{new}}}=-f^{-1} g_{\tilde{z}} b^{-1},
   e_z^{\mbox{\scriptsize{new}}}=-f^{-1} g_{\tilde{w}} b^{-1},
   g_{\tilde{z}}^{\mbox{\scriptsize{new}}}=-b^{-1} e_w f^{-1},
   g_{\tilde{w}}^{\mbox{\scriptsize{new}}}=-b^{-1} e_z f^{-1},\nn
   f^{\mbox{\scriptsize{new}}}=b^{-1},b^{\mbox{\scriptsize{new}}}=f^{-1}.
\end{eqnarray}
The first four equations %in Eq. (\ref{new}) 
can be interpreted as integrability conditions
for the first two equations in (\ref{dYang}). We can easily check
that the last two equations in  (\ref{dYang})
are invariant under this transformation.
%This can be considered as a transformation 
%$\beta: J\rightarrow J^{\mbox{\scriptsize{new}}}$. 

\item $\gamma_0$-transformation (Gilson, Hamanaka and Nimmo (2009)):
\begin{eqnarray}
\left[\begin{array}{cc}
f^{\mbox{\scriptsize{new}}}&g^{\mbox{\scriptsize{new}}}\\
e^{\mbox{\scriptsize{new}}}&b^{\mbox{\scriptsize{new}}}
\end{array}\right]=
\left[\begin{array}{cc}
b&e\\g&f\end{array}\right]^{-1}
=
\left[\begin{array}{cc}
(b-ef^{-1} g)^{-1}&(g-f e^{-1} b)^{-1}\\
(e-b g^{-1} f)^{-1}&(f-g b^{-1} e)^{-1}\end{array}\right].
\label{gamma_0}
\end{eqnarray}
This follows from the fact that
the transformation $\gamma_0: J\mapsto 
J^{\mbox{\scriptsize{new}}}$ is equivalent to 
the simple conjugation
$ J^{\mbox{\scriptsize{new}}}=C_0^{-1}JC_0,~
 C_0=\left[\begin{array}{cc}0&1\\1&0\end{array}\right],$
which clearly leaves the \NC{} Yang equation (\ref{yang}) invariant.
The relation (\ref{gamma_0}) is derived 
by comparing elements in this transformation.
\end{itemize}
It is easy to see that $\beta\circ\beta=
\gamma_0\circ \gamma_0=id$, the identity transformation.

\subsection{Exact \NC{} Atiyah-Ward ansatz solutions} 

Now we construct exact solutions 
by using a chain of B\"acklund transformations from a seed solution.
Let us consider $b=e=f=g=\Delta_0^{-1}$.
%$\epsilon_1 b=\epsilon_1 e
%=\epsilon_2 f=\epsilon_2 g=\Delta_0^{-1}$,
%where $\epsilon^2_s=1$, $s=1,2$.
%(These parameters come from a reflection symmetry of the \NC{} decomposed 
%Yang equation (\ref{dYang}) such as 
%$b\mapsto -b,e\mapsto -e$ and $f\mapsto -f,g\mapsto -g$ and 
%are introduced for discussion in section 4.1 
%about an origin of the present solutions.)
We can easily find that the decomposed \NC{} Yang equation
is reduced to a \NC{} linear equation $(\partial_z\partial_{\tilde{z}}
-\partial_w\partial_{\tilde{w}})\Delta_0=0$. (We note that
for the Euclidean space, this is the \NC{} Laplace equation because of
the reality condition $\bar{w}=-\tilde{w}$.)
Hence we can generate two series of exact solutions
$R_l$ and $R_l^\prime$
by iterating the $\beta$- and $\gamma_0$-transformations
one after the other as follows:
%\begin{eqnarray}
%\begin{array}{ccccccccc}
%R_0&\stackrel{\alpha}{\rightarrow}&R_1&\stackrel{\alpha}{\rightarrow}&
%R_2&\stackrel{\alpha}{\rightarrow}&R_3&\rightarrow&\cdots\\
%&&&&&&&&\\
%{\small{\beta}}\updownarrow~~&{\small{\gamma_0}}\nearrow~~&
%{\small{\beta}}\updownarrow~~&{\small{\gamma_0}}\nearrow~~&
%{\small{\beta}}\updownarrow~~&{\small{\gamma_0}}\nearrow~~&
%{\small{\beta}}\updownarrow~~&\nearrow~~&\cdots\\
%&&&&&&&&\\
%R^\prime_1&\stackrel{\alpha^\prime}{\rightarrow}&
%R^\prime_2&\stackrel{\alpha^\prime}{\rightarrow}&
%R^\prime_3&\stackrel{\alpha^\prime}{\rightarrow}&
%R^\prime_4&\rightarrow&\cdots
%\end{array}
%\end{eqnarray}
\[
\xymatrix{
R_0\ar@{->}[r]^\alpha\ar@{<->}[dr]_\beta&R_1\ar@{->}[r]^\alpha\ar@{<->}[dr]_\beta&R_2\ar@{->}[r]^\alpha\ar@{<->}[dr]_\beta&R_3\ar@{->}[r]^\alpha\ar@{<->}[dr]_\beta&R_4\ar@{->}[r]&\cdots\\
&R'_1\ar@{->}[r]_{\alpha'}\ar@{<->}[u]^{\gamma_0}&R'_2\ar@{->}[r]_{\alpha'}\ar@{<->}[u]^{\gamma_0}&R'_3\ar@{->}[r]_{\alpha'}\ar@{<->}[u]^{\gamma_0}&R'_4\ar@{->}[r]\ar@{<->}[u]^{\gamma_0}&\cdots
}
\]
where $\alpha=\gamma_0
\circ \beta:R_l \rightarrow R_{l+1}$
and $\alpha^{\prime}=\beta \circ \gamma_0: R^\prime_l \rightarrow
R^\prime_{l+1}$. 
%Note that the seed solution 
%$ b= e= f= g=\Delta_0^{-1}$ 
%belongs to $R_0$. 
These two kind of series of solutions 
in fact arise from some class of \NC{} Atiyah-Ward ansatz.
The explicit form of the solutions $R_l$ or $R_l^\prime$
can be represented in terms of quasideterminants
whose elements $\Delta_i$ ($i=-l+1,-l+2,\cdots,l-1$) satisfy
\begin{eqnarray}
\label{chasing}
 \frac{\partial \Delta_i}{\partial z}
= \frac{\partial \Delta_{i+1}}{\partial \tilde{w}},~~~
 \frac{\partial \Delta_i}{\partial w}
= \frac{\partial \Delta_{i+1}}{\partial \tilde{z}},~~~
-l+1\leq i\leq l-2~~~(l\geq 2),
\end{eqnarray}
which imply that every element $\Delta_i$ is a solution of
the \NC{} linear equation $(\partial_z\partial_{\tilde{z}}
-\partial_w\partial_{\tilde{w}})\Delta_i=0$.
A brief introduction of quasideterminants is given in Appendix A.

The results are as follows:
%(given in Gilson, Hamanaka and Nimmo (2009) 
%for the case $==1$) 

\begin{itemize}
 \item \NC{} Atiyah-Ward ansatz solutions $R_l$

The elements in $J_l$ are given explicitly in terms of
quasideterminants of the same $(l+1)\times (l+1)$ matrix:
\begin{eqnarray*}
b_l&=&
%(D_m^{-1})_{mm}=\vert D_m\vert_{mm}^{-1}=
\begin{array}{|cccc|}
\Delta_0&\Delta_{-1} & \cdots & \Delta_{-l}\\
\Delta_1 &\Delta_0&\cdots & \Delta_{1-l} \\
\vdots &\vdots &\ddots & \vdots\\
\Delta_{l} &\Delta_{l-1} &\cdots &\fbox{$\Delta_0$} 
\end{array}^{-1},~~~
f_l=
%(D_m^{-1})_{11}=\vert D_m\vert_{11}^{-1}=
\begin{array}{|cccc|}
\fbox{$\Delta_0$}&\Delta_{-1} & \cdots & \Delta_{-l}\\
\Delta_1 &\Delta_0&\cdots & \Delta_{1-l} \\
\vdots &\vdots &\ddots & \vdots\\
\Delta_{l} &\Delta_{l-1} &\cdots &\Delta_0
\end{array}^{-1},\nonumber\\
e_l&=&
%(D_m^{-1})_{m1}=\vert D_m\vert_{1m}^{-1}=
\begin{array}{|cccc|}
\Delta_0&\Delta_{-1} & \cdots & \fbox{$\Delta_{-l}$}\\
\Delta_1 &\Delta_0&\cdots & \Delta_{1-l} \\
\vdots &\vdots &\ddots & \vdots\\
\Delta_{l} &\Delta_{l-1} &\cdots &\Delta_0
\end{array}^{-1},~~~
g_l=
%&=&(D_m^{-1})_{1m}=\vert D_m\vert_{m1}^{-1}
\begin{array}{|cccc|}
\Delta_0&\Delta_{-1} & \cdots & \Delta_{-l}\\
\Delta_1 &\Delta_0&\cdots & \Delta_{1-l} \\
\vdots &\vdots &\ddots & \vdots\\
\fbox{$\Delta_{l}$} &\Delta_{l-1} &\cdots &\Delta_0 
\end{array}^{-1}.
\end{eqnarray*}
In the commutative limit, 
we can easily check by using \eqref{limit}
that $ b_l=f_l$.
The ansatz $R_0$ leads again 
to the so called the {\it Corrigan-Fairlie-'t Hooft-Wilczek} 
ansatz (Corrigan and Fairlie (1977);
t'Hooft (1976), Wilczek (1977)).
%The relation between index $l$ and the size of the matrices 
%is different from that used in Gilson, Hamanaka and Nimmo (2009).

\item \NC{} Atiyah-Ward ansatz solutions $R^\prime_l$

The elements in $J_l^\prime$ are given explicitly in terms of
quasideterminants of the $l\times l$ matrices:
%\NC{} Atiyah-Ward ansatz solutions $R'_l$
%are represented by the explicit form 
%of elements $b'_l$, $e'_l$, $f'_l$, $g'_l$
%in $J^\prime_l$ as quasideterminants of $l\times l$ matrices:
\begin{eqnarray*}
b^\prime_l
&=&
%(D_l^{-1})_{11}^{-1}=\vert D_l\vert_{11}=
\begin{array}{|cccc|}
\fbox{$\Delta_0$}&\Delta_{-1} & \cdots & \Delta_{1-l}\\
\Delta_1 &\Delta_0&\cdots & \Delta_{2-l} \\
\vdots &\vdots &\ddots & \vdots\\
\Delta_{l-1} &\Delta_{l-2} &\cdots &\Delta_0 
\end{array}~,~~~
f^\prime_l
=
%(D_l^{-1})_{mm}^{-1}=\vert D_l\vert_{mm}=
\begin{array}{|cccc|}
\Delta_0&\Delta_{-1} & \cdots & \Delta_{1-l}\\
\Delta_1 &\Delta_0&\cdots & \Delta_{2-l} \\
\vdots &\vdots &\ddots & \vdots\\
\Delta_{l-1} &\Delta_{l-2} &\cdots &\fbox{$\Delta_0$}
\end{array}~,\nonumber\\
e^\prime_l
&=&
%(D_l^{-1})_{m1}^{-1}=\vert D_l\vert_{1m}=
\begin{array}{|cccc|}
\Delta_{-1}&\Delta_{-2} & \cdots & \fbox{$\Delta_{-l}$}\\
\Delta_0 &\Delta_{-1}&\cdots & \Delta_{1-l} \\
\vdots &\vdots &\ddots & \vdots\\
\Delta_{l-2} &\Delta_{l-3} &\cdots &\Delta_{-1}
\end{array}~,~~~
g^\prime_l=
%(D_l^{-1})_{1m}^{-1}=\vert D_l\vert_{m1}=
\begin{array}{|cccc|}
\Delta_1&\Delta_{0} & \cdots & \Delta_{2-l}\\
\Delta_2 &\Delta_1&\cdots & \Delta_{3-l} \\
\vdots &\vdots &\ddots & \vdots\\
\fbox{$\Delta_{l}$} &\Delta_{l-1} &\cdots &\Delta_1 
\end{array}~.
\end{eqnarray*}
In the commutative case, 
$ b^\prime_l= f^\prime_l$ also holds.
For $l=1$, we get $ b^\prime_1= f^\prime_1
=\Delta_0, e^\prime_1= \Delta_{-1}, 
g^\prime_1= \Delta_1$ 
and then the relation (\ref{chasing}) implies that 
$e^\prime_{1,z}=f^\prime_{1,\tilde{w}},~
e^\prime_{1,w}=f^\prime_{1,\tilde{z}},~
b^\prime_{1,z}= g^\prime_{1, \tilde{w}},~
b^\prime_{1, w}= g^\prime_{1, \tilde{z}},$
and leads to the Corrigan-Fairlie-'t Hooft-Wilczek ansatz
as first pointed out by Yang (1977).

\end{itemize}

The $\gamma_0$-transformation
is proved simply using the \NC{} Jacobi identity (\ref{nc syl}) 
applied to the four corner elements. 
%(We note that quasideterminants are invariant 
%under any permutations of the rows or columns (Gelfand and Retakh (1991)).)
For example,
\begin{eqnarray*}
 b_{l}^{-1}%&=&
%\begin{array}{|cccc|}
%\Delta_0&\Delta_{-1} & \cdots & \Delta_{-l}\\
%\Delta_1 &\Delta_0&\cdots & \Delta_{1-l} \\
%\vdots &\vdots &\ddots & \vdots\\
%\Delta_{m}&\Delta_{m-1} &\cdots &\fbox{$\Delta_0$} 
%\end{array}\nonumber\\
&=&
\begin{array}{|ccc|}
\Delta_0&\cdots & \Delta_{1-l}\\
\vdots &\ddots & \vdots\\
\Delta_{l-1} &\cdots &\fbox{$\Delta_0$} 
\end{array}
-\begin{array}{|ccc|}
\Delta_1&\cdots & \Delta_{2-l}\\
\vdots &\ddots & \vdots\\
\fbox{$\Delta_{l}$} &\cdots &\Delta_1 
\end{array}~
\begin{array}{|ccc|}
\fbox{$\Delta_0$}&\cdots & \Delta_{1-l}\\
\vdots &\ddots & \vdots\\
\Delta_{l-1} &\cdots &\Delta_0 
\end{array}^{-1}
\begin{array}{|ccc|}
\Delta_{-1} &\cdots & \fbox{$\Delta_{-l}$} \\
\vdots &\ddots & \vdots\\
\Delta_{l-2} &\cdots &\Delta_{-1} 
\end{array}
\nonumber\\  
&=&(f^\prime_l-g^\prime_l b^{\prime -1}_l
e^\prime_l).
\end{eqnarray*}

The proof of the $\beta$-transformation
uses both the \NC{} Jacobi identity (\ref{nc syl}) and also 
the homological relations \eqref{row hom}. 
We will consider the first equation in the $\beta$-transformation:
\begin{eqnarray}
 e'_{l,w}=f_{l-1}^{-1}g_{l-1,\tilde z}b_{l-1}^{-1}.
\end{eqnarray}
The RHS is equal to
\begin{eqnarray}
 -b_{l}'g_{l-1}(g_{l-1}^{-1})_{\tilde z}g_{l-1}f_{l}'.
\end{eqnarray}
In this, it follows from \eqref{row hom} 
that the first two and last two factors are
\begin{eqnarray}
b_{l}'g_{l-1}=
\begin{vmatrix}
\fbox{$0$}&\Delta_{-1}&\cdots&\Delta_{1-l}\\
0&\Delta_{0}&\cdots&\Delta_{2-l}\\
\vdots&\vdots&&\vdots\\
0&\Delta_{l-3}&\cdots&\Delta_{-1}\\
1&\Delta_{l-2}&\cdots&\Delta_{0}
\end{vmatrix},~~~
g_{l-1}f_l'=
\begin{vmatrix}
\Delta_{0}&\Delta_{-1}&\cdots &\Delta_{2-l} &\Delta_{1-l}\\
\vdots&\vdots&&\vdots&\vdots\\
\Delta_{l-2}&\Delta_{l-3}&\cdots&\Delta_{0}&\Delta_{-1}\\
1&0&\cdots&0&\fbox{$0$}
\end{vmatrix}.
\end{eqnarray}

Next, from \eqref{col diff}, we have 
\[
\begin{split}
&(g_{l-1}^{-1})_{\tilde z}=
\begin{vmatrix}
\Delta_{0,\tilde z}&\Delta_{-1}&\cdots&\Delta_{1-l}\\
\Delta_{1,\tilde z}&\Delta_{0}&\cdots&\Delta_{2-l}\\
\vdots&\vdots&&\vdots\\
\Delta_{l-2,\tilde z}&\Delta_{l-3}&\cdots&\Delta_{-1}\\
\fbox{$\Delta_{l-1,\tilde z}$}&\Delta_{l-2}&\cdots&\Delta_{0}
\end{vmatrix}\\&+\sum_{k=1}^{l-1}
\begin{vmatrix}
\Delta_{-k,\tilde z}&\Delta_{-1}&\cdots&\Delta_{1-l}\\
\Delta_{1-k,\tilde z}&\Delta_{0}&\cdots&\Delta_{2-l}\\
\vdots&\vdots&&\vdots\\
\Delta_{l-2-k,\tilde z}&\Delta_{l-3}&\cdots&\Delta_{-1}\\
\fbox{$\Delta_{l-1-k,\tilde z}$}&\Delta_{l-2}&\cdots&\Delta_{0}
\end{vmatrix}
\begin{vmatrix}
\Delta_{0}&\Delta_{-1}&\cdots &\Delta_{-k} &\cdots&\Delta_{1-l}\\
\vdots&\vdots&&\vdots&&\vdots\\
\Delta_{l-2}&\Delta_{l-3}&\cdots&\Delta_{l-2-k}&\cdots&\Delta_{-1}\\
\fbox{$0$}&0&\cdots&1&\cdots&0
\end{vmatrix}.
\end{split}
\]
The effect of the left and right factors
on this expression is to move expansion points
as specified in \eqref{row hom}, obtaining 
\[
\begin{split}
&f_{l-1}^{-1}g_{l-1,\tilde z}b_{l-1}^{-1}=
-\begin{vmatrix}
\Delta_{-1}&\cdots&\Delta_{1-l}&\fbox{$\Delta_{1-l,\tilde z}$}\\
\Delta_{0}&\cdots&\Delta_{2-l}&\Delta_{2-l,\tilde z}\\
\vdots&&\vdots&\vdots\\
\Delta_{l-3}&\cdots&\Delta_{-1}&\Delta_{-1,\tilde z}\\
\Delta_{l-2}&\cdots&\Delta_{0}&\Delta_{0,\tilde z}
\end{vmatrix}\\&
-\sum_{k=0}^{l-2}
\begin{vmatrix}
\Delta_{-1}&\cdots&\Delta_{1-l}&\fbox{$\Delta_{-k,\tilde z}$}\\
\Delta_{0}&\cdots&\Delta_{2-l}&\Delta_{1-k,\tilde z}\\
\vdots&&\vdots&\vdots\\
\Delta_{l-2}&\cdots&\Delta_{0}&\Delta_{l-1-k,\tilde z}
\end{vmatrix}
\begin{vmatrix}
0&\cdots&1&\cdots&0&\fbox{$0$}\\
\Delta_{0}&\cdots &\Delta_{-k} &\cdots&\Delta_{2-l}&\Delta_{1-l}\\
\vdots&&\vdots&&\vdots&\vdots\\
\Delta_{l-2}&\cdots&\Delta_{l-2-k}&\cdots&\Delta_{0}&\Delta_{-1}
\end{vmatrix}.
\end{split}
\]
On the other hand, 
\[
\begin{split}
& e'_{l,w}=
\begin{vmatrix}
\Delta_{-1}&\cdots&\Delta_{1-l}&\fbox{$\Delta_{-l,w}$}\\
\Delta_{0}&\cdots&\Delta_{2-l}&\Delta_{1-l,w}\\
\vdots&&\vdots&\vdots\\
\Delta_{l-3}&\cdots&\Delta_{-1}&\Delta_{-2,w}\\
\Delta_{l-2}&\cdots&\Delta_{0}&\Delta_{-1,w}
\end{vmatrix}\\&+\sum_{k=0}^{l-2}
\begin{vmatrix}
\Delta_{-1}&\cdots&\Delta_{1-l}&\fbox{$\Delta_{-k-1,w}$}\\
\Delta_{0}&\cdots&\Delta_{2-l}&\Delta_{-k,w}\\
\vdots&&\vdots&\vdots\\
\Delta_{l-2}&\cdots&\Delta_{0}&\Delta_{l-2-k,w}
\end{vmatrix}
\begin{vmatrix}
0&\cdots&1&\cdots&0&\fbox{$0$}\\
\Delta_{0}&\cdots &\Delta_{-k} &\cdots&\Delta_{2-l}&\Delta_{1-l}\\
\vdots&&\vdots&&\vdots&\vdots\\
\Delta_{l-2}&\cdots&\Delta_{l-2-k}&\cdots&\Delta_{0}&\Delta_{-1}
\end{vmatrix},
\end{split}
\]
and then the result follows immediately 
from $\Delta_{i,w}=\Delta_{i+1,\tilde z}$ in \eqref{chasing}.

We can find that the proof of these results relies on
using quasideterminant identities alone. Thus we can conclude that
{\it \NC{} B\"acklund transformations are identities of quasideterminants}.
%This is a natural analogue in lower-dimensional 
%commutative integrable systems. (e.g. Hirota (2004).)

\bigskip

We can also present a compact form of the whole of Yang's $J$-matrix 
in terms of a single quasideterminant expanded by a $2\times2$
submatrix:
\begin{align*}
\begin{vmatrix}
a&b&c\\
d&\fbox{$e$}&\fbox{$f$}\\
g&\fbox{$h$}&\fbox{$i$}\\
\end{vmatrix}&
:=
\begin{bmatrix}
\begin{vmatrix}
a&b\\
d&\fbox{$e$}
\end{vmatrix}&\begin{vmatrix}
a&c\\
d&\fbox{$f$}
\end{vmatrix}
\\
\begin{vmatrix}
a&b\\
g&\fbox{$h$}\\
\end{vmatrix}&
\begin{vmatrix}
a&c\\
g&\fbox{$i$}\\
\end{vmatrix}
\end{bmatrix}.
\end{align*}
%The results are as follows.
The solutions for the $J$-matrix can be presented as follows:

\begin{itemize}

 \item \NC{} Atiyah-Ward ansatz solutions $R_l$

\begin{eqnarray*}
\label{J}
J_l=
%\left[
%\begin{array}{cc}
%f'_l-g'_lb_l^{\prime-1}e'_l&-g'_lb_l^{\prime-1}\\
%b_l^{\prime-1}e'_l&b_l^{\prime-1}
%\end{array}\right]=
\begin{pmat}|{||..|}|
\fbox{0}&-1&0&\cdots&0&\fbox{0}\cr\-
1&\Delta_0&\Delta_{-1}&\cdots&\Delta_{1-l}&\Delta_{-l}\cr\-
%\Delta_2&\Delta_1&\Delta_0&\cdots&\Delta_{-l+4}&\Delta_{-l+3}&0\cr
0&\Delta_1&\Delta_0&\cdots&\Delta_{2-l}&\Delta_{1-l}
\cr
\vdots&\vdots&\vdots&\ddots&\vdots&\vdots\cr
0&\Delta_{l-1}&\Delta_{l-2}&\cdots&\Delta_{0}&\Delta_{-1}
\cr\-
\fbox{0}&\Delta_{l}&\Delta_{l-1}&\cdots&\Delta_1&\fbox{$\Delta_{0}$}\cr
\end{pmat},~
J_l^{-1}=
%\left[
%\begin{array}{cc}
%f'_l-g'_lb_l^{\prime-1}e'_l&-g'_lb_l^{\prime-1}\\
%b_l^{\prime-1}e'_l&b_l^{\prime-1}
%\end{array}\right]=
\begin{pmat}|{|..||}|
\fbox{$\Delta_0$}&\Delta_{-1}&\cdots&\Delta_{1-l}&\Delta_{-l}&\fbox{0}\cr\-
%\Delta_2&\Delta_1&\Delta_0&\cdots&\Delta_{-l+4}&\Delta_{-l+3}&0\cr
\Delta_1&\Delta_0&\cdots&\Delta_{2-l}&\Delta_{1-l}&0\cr
\vdots&\vdots&\ddots&\vdots&\vdots&\vdots\cr
\Delta_{l-1}&\Delta_{l-2}&\cdots&\Delta_{0}&\Delta_{-1}&0
\cr\-
\Delta_{l}&\Delta_{l-1}&\cdots&\Delta_{1}&\Delta_{0}&1\cr\-
\fbox{0}&0&\cdots&0&-1&\fbox{0}\cr
\end{pmat}.
\end{eqnarray*}

 \item \NC{} Atiyah-Ward ansatz solutions $R_l^\prime$

\begin{eqnarray*}
\label{J_q}
J_l^{\prime}=
%\left[
%\begin{array}{cc}
%f'_l-g'_lb_l^{\prime-1}e'_l&-g'_lb_l^{\prime-1}\\
%b_l^{\prime-1}e'_l&b_l^{\prime-1}
%\end{array}\right]=
\begin{pmat}|{|..||}|
\Delta_0&\Delta_{-1}&\cdots&\Delta_{1-l}&\Delta_{-l}&-1\cr\-
\Delta_1&\Delta_0&\cdots&\Delta_{2-l}&\Delta_{1-l}&0\cr
%\Delta_2&\Delta_1&\Delta_0&\cdots&\Delta_{-l+4}&\Delta_{-l+3}&0\cr
\vdots&\vdots&\ddots&\vdots&\vdots&\vdots\cr
\Delta_{l-1}&\Delta_{l-2}&\cdots&\Delta_0&\Delta_{-1}&0\cr\-
\Delta_{l}&\Delta_{l-1}&\cdots&\Delta_1&\fbox{$\Delta_{0}$}&\fbox{0}\cr\-
1&0&\cdots&0&\fbox{0}&\fbox{0}\cr
\end{pmat},~
%\label{J^-1_q}
J_l^{\prime -1}=
%\left[
%\begin{array}{cc}
%f_l^{\prime-1}&f_l^{\prime-1}g'_l\\
%-e'_lf_l^{\prime-1}&b'_l-e'_lf_l^{\prime-1}g'_l
%\end{array}\right]=
\begin{pmat}|{||..|}|
\fbox{0}&\fbox{0}&0&\cdots&0&1\cr\-
\fbox{0}&\fbox{$\Delta_0$}&\Delta_{-1}&\cdots&\Delta_{1-l}&\Delta_{-l}\cr\-
0&\Delta_1&\Delta_0&\cdots&\Delta_{2-l}&\Delta_{1-l}\cr
\vdots&\vdots&\vdots&\ddots&\vdots&\vdots\cr
0&\Delta_{l-1}&\Delta_{l-2}&\cdots&\Delta_0&\Delta_{-1}\cr\-
-1&\Delta_{l}&\Delta_{l-1}&\cdots&\Delta_1&\Delta_{0}\cr
\end{pmat}.
\end{eqnarray*}

\end{itemize}

%Here lower case letters denote single entries, upper case letters
%denote matrices of compatible dimensions and Greek letters are
%scalars (i.e.{} commute with everything). 
Because $J$ is gauge invariant, this shows that
the present B\"acklund transformation is not
just a gauge transformation but a nontrivial transformation.

The proof of these representations 
is given by using the \NC{} Jacobi identity, homological relations
and the inversion formula for $J$:
\begin{eqnarray}
 J^{-1}=
\left[\begin{array}{cc} f^{-1} &f^{-1} g \\
-e f^{-1} & b-e f^{-1} g
   \end{array}
   \right],
\end{eqnarray}
or simply using the formula \eqref{inverse}. 
(For a detailed proof, see Appendix A in 
Gilson, Hamanaka and Nimmo (2009).)

\subsection{Some Explicit Examples}

By solving the \NC{} linear equation 
$(\partial_z\partial_{\tilde{z}}
-\partial_w\partial_{\tilde{w}})\Delta_0=0$
for the seed solution of the B\"acklund transformations, 
we can obtain exact solutions explicitly.

For example, in Euclidean space, the \NC{} linear
equation is just the 4-dimensional \NC{} Laplace equation
whose solutions include a \NC{} version of the fundamental solution:
$\Delta_0=1+\sum_{p=1}^{k}(a_p/(z\tilde{z}-w\tilde{w}))$ 
($a_p$ are constants), which leads to \NC{} instanton solutions whose instanton
number is $k$ (Correa, Lozano, Moreno and Schaposnik (2001);
Lechtenfeld and  Popov (2002); Nekrasov and Schwarz (1998)). 
The B\"acklund transformations do not increase the instanton number.

There is also a simple new solution:
\begin{eqnarray}
 \Delta_0=c\exp(az+b\tilde{z}+aw+b\tilde{w}),
\end{eqnarray}
where $a,b$ and $c$ are constants.
This leads to a \NC{} version of 
nonlinear plane wave solutions (de Vega (1988)).
These solutions behave as standard solitons in lower-dimension
and do not decay at infinity, which implies that
this gives an infinite value for the Yang-Mills action.
By following the analysis used in Hamanaka (2007) for the \NC{} KP equation, the asymptotic behaviour of these solutions can be shown to be the same as the corresponding commutative ones.

Other solutions are also easily obtained and a more detailed discussion on this topic
will be reported elsewhere.

\section{Twistor descriptions of the \NC{} ASDYM equations}

In this section, we explain the origin of the 
B\"acklund transformations for the \NC{} ASDYM equations
and \NC{} Atiyah-Ward ansatz solutions from the geometrical
viewpoint of \NC{} twistor theory. Here we just need a one-to-one 
correspondence between a solution of the \NC{} ASDYM equations and 
a patching matrix $P=P(\zeta w+\zt,\zeta z+\wt,\zeta)$ 
of a \NC{} holomorphic vector bundle 
on a \NC{} 3-dimensional projective space, which is
called the \NC{} Penrose-Ward correspondence.
This correspondence is established in the Moyal-deformed case
by Brain (2005); Brain and Majid (2008);
Kapustin, Kuznetsov and Orlov (2001);
Lechtenfeld and  Popov (2002); Takasaki (2001)
and here we apply their formal procedure to general noncommutative situations.
Such twistor treatments are useful not only for 
constructing exact solutions but also 
for checking whether the B\"acklund transformation 
act on the solution spaces transitively.

In order to review this correspondence briefly, 
we introduce another linear system defined on
another local patch whose (commutative) 
coordinate is $\tilde{\zeta}=1/\zeta$,
\begin{eqnarray}
(\tilde{\zeta} D_w -D_{\tilde{z}})\tilde{\psi}=0,\nn
(\tilde{\zeta} D_z-D_{\tilde{w}})\tilde{\psi}= 0.
\label{lin_asdym2}
\end{eqnarray}
A nontrivial solution $\tilde{\psi}$ ($N\times N$ matrix)
of the linear system (\ref{lin_asdym2}) 
is supposed to exist and 
is not regular at $\tilde{\zeta}=\infty$ (or equivalently $\zeta=0$)
as discussed earlier for $\psi$.
%(The existence is supposed.)

Any solution of the \NC{} ASDYM equation determines 
solution $\psi$ and $\tilde{\psi}$ that are unique
up to gauge transformation, and then
the corresponding patching matrix is given by
\begin{eqnarray}
%\label{birkhoff}
 P(x;\zeta)=\tilde{\psi}^{-1}(x;\zeta)\psi(x;\zeta).
% \psi=\tilde{\psi}P,
\end{eqnarray}

Conversely, if a patching matrix $P=P(\zeta w+\zt,\zeta z+\wt,\zeta)$ 
is factorized as 
\begin{eqnarray}
%\label{birkhoff}
 P(\zeta w+\zt,\zeta z+\wt,\zeta)=\tilde{\psi}^{-1}(x;\zeta)\psi(x;\zeta),
% \psi=\tilde{\psi}P,
\end{eqnarray}
where $\psi$ and $\tilde{\psi}$ are regular near $\zeta=0$
and $\zeta=\infty$, respectively, 
then $\psi$ and $\tilde{\psi}$ are solutions 
of the linear system (\ref{lin_asdym}) for the \NC{} ASDYM equation.
Then we can recover the ASDYM gauge fields 
in terms of $h$ and $\tilde{h}$ by using
\eqref{a} and the fact that
$h(x)=\psi(x,\zeta=0), \tilde{h}(x)=\tilde{\psi}(x,\zeta=\infty)$.
(We can easily understand this by comparing the linear systems
\eqref{lin_asdym} and  \eqref{lin_asdym2} with \eqref{h}.)

In the commutative case, if $P$ is holomorphic w.r.t. $\zeta$,
then the factorization is guaranteed by the Birkhoff factorization theorem.
In the case of the Moyal deformation, this is formally proved by Takasaki (2001).
%(We note that $\zeta$ commute with all variables.)
Here we will see that under the Atiyah-Ward ansatz for
the patching matrix, the factorization problem 
(the {\it Riemann-Hilbert problem}) is solved.

\bigskip

In this section, we fix the gauge to be, what we call in this paper, the {\it Mason-Woodhouse gauge}
%for simplicity which realize the parameterization of $J$:
\begin{eqnarray}
 J 
=\left[\begin{array}{cc} f -g b^{-1} e&-g b^{-1}
   \\ b^{-1}  e &b^{-1}\end{array}
   \right]
=\left[\begin{array}{cc}1&g
   \\ 0 &b\end{array}
   \right]^{-1}
 \left[\begin{array}{cc}f&0
   \\ e&1\end{array}
   \right]
= \tilde{h}_{\scr\mbox{MW}}^{-1} h_{\scr\mbox{MW}}.%,\nn
%\end{eqnarray}
%Then the inverse of $J$ can be easily calculated as 
%\begin{eqnarray} 
\label{M-W}
% J^{-1}=h^{-1}\tilde{h}=
%   \left[\begin{array}{cc}f&0
%   \\ e &1\end{array}
%   \right]^{-1}
% \left[\begin{array}{cc}1&g
%   \\ 0&b\end{array}
%   \right]
%=
%\left[\begin{array}{cc} f^{-1} &f^{-1} g \\
%-e f^{-1} & b-e f^{-1} g
%   \end{array}
%   \right].
\end{eqnarray}
We note that the gauge transformation $g={\mbox{diag}}(f^{1/2},b^{1/2})$
connects the Mason-Woodhouse gauge with a \NC{} version of Yang's
$R$-gauge (Yang (1977)):
\begin{eqnarray}
 J 
=\left[\begin{array}{cc} f -g b^{-1} e&-g b^{-1}
   \\ b^{-1}  e &b^{-1}\end{array}
   \right]
=\left[\begin{array}{cc}f^{-1/2}&f^{-1/2}g
   \\ 0 &b^{1/2}\end{array}
   \right]^{-1}
 \left[\begin{array}{cc}f^{1/2}&0
   \\ b^{-1/2}e&b^{-1/2}\end{array}
   \right]
= \tilde{h}_{\scr\mbox{R}}^{-1} h_{\scr\mbox{R}},%,\nn
\end{eqnarray}
where the square root is considered as
any quantity which satisfies
$f^{1/2}f^{1/2}=f,~
f^{-1/2}:=(f^{1/2})^{-1}$ and whenever this notation is used, it is assumed to exist.

% Since $h$ and $\tilde h$ are known from \eqref{M-W}, 
%the corresponding gauge fields can be calculated explicitly from \eqref{a}.

The wave functions $\psi$ and $\tilde{\psi}$ can be
expanded by $\zeta$ and $\tilde{\zeta}=1/\zeta$, respectively:
\begin{eqnarray}
\label{expansion}
\psi&=&h+{\cal{O}}(\zeta)
=\left[\begin{array}{cc}
h_{11}+\sum_{i=1}^{\infty}a_i\zeta^i&
h_{12}+\sum_{i=1}^{\infty}b_i\zeta^i\\
h_{21}+\sum_{i=1}^{\infty}c_i\zeta^i&
h_{22}+\sum_{i=1}^{\infty}d_i\zeta^i.
\end{array}\right],\nn
\tilde{\psi}&=&\tilde{h}+{\cal{O}}(\tilde{\zeta})
=
\left[\begin{array}{cc}
\tilde{h}_{11}+\sum_{i=1}^{\infty}\at_i\tilde{\zeta}^i&
\tilde{h}_{12}+\sum_{i=1}^{\infty}\bt_i\tilde{\zeta}^i\\
\tilde{h}_{21}+\sum_{i=1}^{\infty}\ct_i\tilde{\zeta}^i&
\tilde{h}_{22}+\sum_{i=1}^{\infty}\dt_i\tilde{\zeta}^i.
\end{array}\right].
\end{eqnarray}

%\subsection{Brief Introduction to Twistor Theory}

\subsection{Riemann-Hilbert problem for \NC{} Atiyah-Ward Ansatz}

%This patching matrix is in fact holomorphic in some sense,
%and by Birkhoff factorization, we have for a fixed
%space coordinates $(z,w,\zt,\wt)$,
%\begin{eqnarray}
% P(\zeta w+\zt,\zeta z+\wt,\zeta)=\tilde{\psi}^{-1}\psi.
%\end{eqnarray}
%Let us note that we take the following gauge
%in the \NC{} Corrigan-Fairlie-Yates-Goddard transformation:
%\begin{eqnarray}
%\label{normalization}
% &&\psi(x,\zeta=0)=h(x)=
%\left[\begin{array}{cc}f&0
%            \\ e &1\end{array}
%   \right],\\
%&&\tilde{\psi}(x,\zeta=\infty)=\tilde{h}(x)
%=
%\left[\begin{array}{cc}1&g
%      \\ 0&b\end{array}
%   \right].
%\end{eqnarray}
%{\sf There are many things to be added.}

{}From now on, we restrict ourselves to $G=GL(2)$.
In this case, we can take a simple ansatz for
the patching matrix $P$, which is called the 
{Atiyah-Ward} ansatz in the commutative case (Atiyah and Ward (1978)). The
\NC{} generalization of this ansatz is straightforward 
and actually leads to a solution of the factorization problem.
The $l$-th order \NC{} Atiyah-Ward ansatz ($l=0,1,2,\dots$)
is specified by choosing the patching matrix
to be:
\begin{eqnarray}
 P_l(x;\zeta)=\left[\begin{array}{cc}0&\zeta^{-l}  
      \\\zeta^{l} &\Delta(x;\zeta)\end{array}
   \right].
\end{eqnarray}
(The standard representation of the ansatz is not $P_l$
but $C_0 P_l$. Both representations are essentially the same.)
We note that the coordinate dependence of 
$P_l=P_l(\zeta w+\zt,\zeta z+\wt,\zeta)$ implies that
$(\del_w-\zeta\del_{\zt})\Delta=0,~(\del_z-\zeta
\del_{\wt}) \Delta=0$. Hence, the Laurent expansion of $\Delta$ 
w.r.t. $\zeta$
\begin{eqnarray}
 \Delta(x;\zeta) = \sum_{i=-\infty}^{\infty}\Delta_i(x) \zeta^{-i},
\end{eqnarray}
gives rise to the following relationships amongst the coefficients,
\begin{eqnarray}
 \frac{\partial \Delta_i}{\partial z}
= \frac{\partial \Delta_{i+1}}{\partial \tilde{w}},~~~
 \frac{\partial \Delta_i}{\partial w}
= \frac{\partial \Delta_{i+1}}{\partial \tilde{z}},
\end{eqnarray}
which coincide with the recurrence relation \eqref{chasing}.
We will soon see that the coefficients $\Delta_i(x)$
are the scalar functions in the solutions generated by
the B\"acklund transformations in the previous section.

We will now solve the factorization problem $\tilde{\psi}P_l= \psi$
for the \NC{} Atiyah-Ward ansatz. In explicit form this is
\begin{eqnarray}
\left[\begin{array}{cc} \tilde{\psi}_{11}&\tilde{\psi}_{12}
            \\ \tilde{\psi}_{21}&\tilde{\psi}_{22}\end{array}
   \right]
\left[\begin{array}{cc}0&\zeta^{-l} 
      \\ \zeta^l & \Delta(x;\zeta)\end{array}
   \right]
=
\left[\begin{array}{cc}\psi_{11}&\psi_{12}
      \\ \psi_{21}&\psi_{22}\end{array}
   \right],
\end{eqnarray}
where $\psi_{ij}$ is the $(i,j)$-th element of $\psi$, and so,
\begin{eqnarray}
\label{splitting1}
&& \tilde{\psi}_{12}\zeta^l=\psi_{11},~~~
\tilde{\psi}_{22}\zeta^l=\psi_{21},\\
&&\tilde{\psi}_{11} \zeta^{-l}
+ \tilde{\psi}_{12}\Delta 
=\psi_{12},~~~
\tilde{\psi}_{21} \zeta^{-l} + \tilde{\psi}_{22}\Delta
=\psi_{22}.
\label{splitting2}
\end{eqnarray}
{}From \eqref{expansion} and \eqref{splitting1},
%the relation between $\psi, \tilde{\psi}$ and $h, \tilde{h}$ (), 
we find that some entries in $\psi$ and $\tilde\psi$ are polynomials w.r.t. $\zeta$
and ${\tilde{\zeta}}=\zeta^{-1}$:
\begin{eqnarray}
 \psi_{11}&=&h_{11}+a_1\zeta+a_2\zeta^2+\cdots +a_{l-1}\zeta^{l-1}
+\tilde{h}_{12}\zeta^l,\nn
 \psi_{21}&=&h_{21}+c_1\zeta+c_2\zeta^2+\cdots +c_{l-1}\zeta^{l-1}
+\tilde{h}_{22}\zeta^l,\nn
 \tilde{\psi}_{12}&=&\tilde{h}_{12}+a_{l-1}\zeta^{-1}
+a_{l-2}\zeta^{-2}+\cdots +a_{1}\zeta^{1-l}
+h_{11} \zeta^{-l},\nn
 \tilde{\psi}_{22}&=&\tilde{h}_{22}+ c_{l-1}\zeta^{-1}
+c_{l-2}\zeta^{-2}+\cdots+c_1 \zeta^{1-l}
+h_{21} \zeta^{-l}.
\end{eqnarray}
By substituting these formulae into \eqref{splitting2},
we get the following sets of equations for $h$ and $\tilde{h}$ in the coefficients
of $\zeta^{0}, ~\zeta^{-1},~ \cdots, ~\zeta^{-l}$:
\begin{eqnarray}
(h_{11}, a_1,\cdots,a_{l-1},\tilde{h}_{12})
D_{l+1}=(-\tilde{h}_{11},0,\cdots,0,h_{12}),\nn
(h_{21}, c_1,\cdots,c_{l-1},\tilde{h}_{22})
D_{l+1}=(-\tilde{h}_{21},0,\cdots,0,h_{22}),
\end{eqnarray}
where
\begin{eqnarray}
 D_l:=
\left[
\begin{array}{cccc}
\Delta_0&\Delta_{-1} & \cdots & \Delta_{1-l}\\
\Delta_1 &\Delta_0&\cdots & \Delta_{2-l} \\
\vdots &\vdots &\ddots & \vdots\\
\Delta_{l-1} &\Delta_{l-2} &\cdots &\Delta_0 
\end{array}
\right].
\end{eqnarray}
These \NC{} linear equations can be solved 
in terms of quasideterminants (cf. \eqref{def_q}) as
\begin{eqnarray}
\label{sol_birkhoff}
 h_{11}&=&h_{12}\vert D_{l+1}\vert_{1,l+1}^{-1}
        -\tilde{h}_{11} \vert D_{l+1}\vert_{1,1}^{-1},\nn
 h_{21}&=&h_{22}\vert D_{l+1}\vert_{1,l+1}^{-1}
        -\tilde{h}_{21} \vert D_{l+1}\vert_{1,1}^{-1},\nn
 \tilde{h}_{12}&=&h_{12}\vert D_{l+1}\vert_{l+1,l+1}^{-1}
        -\tilde{h}_{11} \vert D_{l+1}\vert_{l+1,1}^{-1},\nn
 \tilde{h}_{22}&=&h_{22}\vert D_{l+1}\vert_{l+1,l+1}^{-1}
        -\tilde{h}_{21} \vert D_{l+1}\vert_{l+1,1}^{-1}.
\end{eqnarray}
In \eqref{sol_birkhoff} there are four equations but eight unknowns and so in order to solve them it is necessary to impose four conditions corresponding to a choice of gauge. In particular, the Mason-Woodhouse gauge 
($h_{12}=\tilde{h}_{21}=0$, $\tilde{h}_{11}=h_{22}=1$)
leads to simple representations:
\begin{eqnarray}
h_{11}&=&-
%(D_l^{-1})_{11}=\vert D_l\vert_{11}^{-1}=
\begin{array}{|cccc|}
\fbox{$\Delta_0$}&\Delta_{-1} & \cdots & \Delta_{-l}\\
\Delta_1 &\Delta_0&\cdots & \Delta_{1-l} \\
\vdots &\vdots &\ddots & \vdots\\
\Delta_{l} &\Delta_{l-1} &\cdots &\Delta_0
\end{array}^{-1},~~~
h_{21}=%(D_l^{-1})_{m1}=\vert D_l\vert_{1m}^{-1}=
\begin{array}{|cccc|}
\Delta_0&\Delta_{-1} & \cdots & \fbox{$\Delta_{-l}$}\\
\Delta_1 &\Delta_0&\cdots & \Delta_{1-l} \\
\vdots &\vdots &\ddots & \vdots\\
\Delta_{l} &\Delta_{l-1} &\cdots &\Delta_0
\end{array}^{-1},
\nonumber\\
\tilde{h}_{12}%&=&(D_l^{-1})_{1m}=\vert D_l\vert_{m1}^{-1}
&=&-
\begin{array}{|cccc|}
\Delta_0&\Delta_{-1} & \cdots & \Delta_{-l}\\
\Delta_1 &\Delta_0&\cdots & \Delta_{1-l} \\
\vdots &\vdots &\ddots & \vdots\\
\fbox{$\Delta_{l}$} &\Delta_{l-2} &\cdots &\Delta_0 
\end{array}^{-1},~~~
\tilde{h}_{22}=%(D_l^{-1})_{mm}=\vert D_l\vert_{mm}^{-1}=
\begin{array}{|cccc|}
\Delta_0&\Delta_{-1} & \cdots & \Delta_{-l}\\
\Delta_1 &\Delta_0&\cdots & \Delta_{1-l} \\
\vdots &\vdots &\ddots & \vdots\\
\Delta_{l} &\Delta_{l-1} &\cdots &\fbox{$\Delta_0$} 
\end{array}^{-1},
\end{eqnarray}
which coincides exactly with the solutions $R_{l}$  generated 
by the B\"acklund transformation in the previous section
except for signs in $f_l$ and $g_l$.
(The mismatch of the signs is not essential because it can be absorbed into
the reflection symmetry $f\mapsto -f,g\mapsto -g$ 
of the \NC{} Yang equation \eqref{dYang}.)  
%with $\epsilon_1=-\epsilon_2=1$.
That is why we call them the \NC{} Atiyah-Ward ansatz solutions.
The class of solutions $R_l^\prime$ is also obtained 
in the same way by starting with the Atiyah-Ward ansatz
$C_0^{-1}P_lC_0$.

\subsection{Origin of the \NC{} Corrigan-Fairlie-Yates-Goddard transformation}

Finally let us discuss the origin of the noncommutative Corrigan-Fairlie-Yates-Goddard
transformation, constructed from the $\beta$-transformation and
the $\gamma_0$-transformation and give a generalization of it.
%Finally let us discuss the origin of 
%the \NC{} Corrigan-Fairlie-Yates-Goddard transformations, $\beta$-transformation 
%and $\gamma_0$-transformation and give a generalization of it.
Such geometrical understanding is useful when discussing  whether the
group action of the B\"acklund transformations 
is transitive and hence to find the symmetry of the noncommutative
ASDYM equation.
%Such geometrical understanding is useful to 
%discuss whether the group action of the B\"acklund transformations 
%is transitive in order to find the symmetry of \NC{} ASDYM equation.
The present results are essentially due to Mason, Chakravarty and
Newman (1988), Mason and Woodhouse (1996).
These transformations can be viewed as adjoint actions
of the patching matrix $P$:
\begin{eqnarray}
 \beta: P \mapsto P^{\mbox{\scr{new}}} = B^{-1} P B,~~~
 \gamma_0: P \mapsto P^{\mbox{\scr{new}}} = C_0^{-1} P  C_0,
\end{eqnarray}
where
\begin{eqnarray}
 B=\left[\begin{array}{cc}0&1\\\zeta^{-1}&0\end{array}\right],~~~
 C_0=\left[\begin{array}{cc}0&1\\1&0\end{array}\right].
\end{eqnarray}
It is obvious that $\beta\circ \beta=id, \gamma_0\circ \gamma_0=id$.
%together with a singular gauge transformation for
%$\beta$-transformation. 

The composition of these transformations actually 
maps the $l$-th Atiyah-Ward ansatz to the $(l+1)$-th one:
\begin{eqnarray}
P_l\mapsto 
C_0^{-1}B^{-1}
\left[\begin{array}{cc}0&\zeta^{-l} 
      \\ \zeta^l & \Delta\end{array}
   \right]BC_0
 =
\left[\begin{array}{cc}0&\zeta^{-(l+1)} 
      \\ \zeta^{l+1} & \Delta\end{array}
   \right]=P_{l+1}.
\end{eqnarray}

The action of $C_0$ leads to $h\mapsto hC_0, \tilde{h}\mapsto
\tilde{h}C_0$ and hence to the $\gamma_0$-transformation, and the action of $B$ is defined at the level of 
$\psi$ and $\tilde{\psi}$ 
%$\beta: \psi \mapsto \psi^{\mbox{\scr{new}}}$ and
%$\tilde{\psi} \mapsto \tilde{\psi}^{\mbox{\scr{new}}}$ 
as follows:
\begin{eqnarray}
 \psi^{\mbox{\scr{new}}}=g^{-1}\psi B,~~~
 \tilde{\psi}^{\mbox{\scr{new}}}=g^{-1}\tilde{\psi} B,
\end{eqnarray}
where
\begin{eqnarray}
 g^{-1}=\left[\begin{array}{cc}0&\zeta b^{-1} 
            \\ f^{-1} &0\end{array}
   \right].
\end{eqnarray}
The gauge transformation $g$ is needed to maintain the regularity of $\psi$ and $\tilde{\psi}$, w.r.t $\zeta$ and $\tilde\zeta$ respectively, in the factorization of $P$.

The explicit calculation gives
\begin{eqnarray}
  \psi^{\mbox{\scr{new}}}
= \left[\begin{array}{cc} b^{-1} \psi_{22}& \zeta b^{-1}\psi_{21} 
            \\ \zeta^{-1}f^{-1}\psi_{12} & f^{-1}\psi_{11}
   \end{array}
   \right].
\end{eqnarray}
In the $\zeta\rightarrow 0$ limit, this reduces to
\begin{eqnarray}
  h^{\mbox{\scr{new}}}
=
\left[\begin{array}{cc}f^{\mbox{\scr{new}}}&0
            \\ e^{\mbox{\scr{new}}} &1\end{array}
   \right]=\left[\begin{array}{cc} b^{-1} & 0 
            \\ f^{-1} k_{12} & 1
   \end{array}
   \right],
   \label{beta-trfed}
\end{eqnarray}
where $\psi=h+k\zeta+{\cal{O}}(\zeta^2)$.

Here we note that the linear system (\ref{lin_asdym})  can be
represented in terms of $b,f,e,g$ as
\begin{eqnarray}
 \label{orig}
 L\psi&=&(\del_w-\zeta \del_{\zt})\psi+
   \left[\begin{array}{cc}-f_w f^{-1}&\zeta g_{\zt}b^{-1}
   \\ -e_wf^{-1}&\zeta b_{\zt}b^{-1}\end{array}
   \right]\psi=0,\nn
 M\psi&=&(\del_z-\zeta \del_{\wt})\psi+
   \left[\begin{array}{cc}- f_z f^{-1}&\zeta g_{\wt}b^{-1}
   \\ -e_zf^{-1}&\zeta b_{\wt}b^{-1}\end{array}
   \right]\psi=0.
\end{eqnarray}
By considering the first order term of $\zeta$
in the (1,2) component of the first equation, we find that
\begin{eqnarray}
 \del_w(f^{-1} k_{12})=-f^{-1}g_{\zt}b^{-1}.
\end{eqnarray}
Hence from the (1,1) and (2,1) components of (\ref{beta-trfed}),
we have
\begin{eqnarray}
 f^{\mbox{\scr{new}}}=b^{-1},~~~
 \del_w e^{\mbox{\scr{new}}}=\del_w(f^{-1}k_{12})=-f^{-1}g_{\zt} b^{-1},
\end{eqnarray}
which are just parts of the $\beta$-transformation (\ref{new}).
In similar way, we can get the other ones.
Therefore the $\beta$-transformation (\ref{new})
can be interpreted as the transformation of the
patching matrix $P\mapsto B^{-1} P B$ together with the gauge
transformation $g$.
The results presented in the section and the previous
one lead to a simpler proof of the results in Section 3.
%The present results together with those of the previous section
%give another, simpler proof of section 3.

We note that the $\gamma_0$-transformation can be
generalized to the following transformation (the {\it $\gamma$-transformation}):
\begin{eqnarray}
 \gamma: P \mapsto P^{\mbox{\scr{new}}} = C^{-1} P  C,
\label{gamma}
\end{eqnarray}
where $C$ is an arbitrary constant matrix.
The actions of $\beta$- and $\gamma$- transformations
generate the action of the loop group $LGL(2)$ on $P$
by conjugation.
Therefore the symmetry group of the \NC{} ASDYM equation 
includes the loop group $LGL(2)$ as a subgroup.

\section{Conclusion and Discussion}

In this paper, we have presented B\"acklund transformations for the
\NC{} ASDYM equation with $G=GL(2)$ and constructed from a simple seed solution a
series of exact \NC{} Atiyah-Ward ansatz solutions expressed explicitly
in terms of quasideterminants. We have found that
the B\"acklund transformations generate a wide class of new solutions. 
We have also given the origin of 
the B\"acklund transformation and the generated solutions 
in the framework of \NC{} twistor theory and generalized them. 
%\NC{} Atiyah-Ward ansatz
%works well to solve the \NC{} Riemann-Hilbert problem
%and in some gauge the B\"acklund transformation 

The present results could be taken as the starting point 
to reveal an infinite-dimensional symmetry of the \NC{} ASDYM equation
in terms of some infinite-dimensional algebra. 
We have to prove that the Atiyah-Ward ansatz covers all solutions
of \NC{} ASDYM equation and generalize the B\"acklund transformations
$\beta$ and $\gamma$ so that they should act on the solution
space transitively. 
%From this one might classify 
%lower-dimensional \NC{} integrable systems via reductions, 
%and analyze the behaviour of exact solutions to apply them 
%to the corresponding physical situations. 
%We have found that the quasideterminant formulation is suitable for  
%a \NC{} extension of 4-dimensional ASDYM equation 
%as well as lower-dimensional \NC{} integrable systems such as the KdV eq. 
%This is partly because in both situations solving the \NC{} equations
%reduces to a \NC{} Riemann-Hilbert problem. Then the
%original problem becomes that of linear algebra and matrix inversion.
%Thus it is natural that quasideterminants appear.
%Then original problem becomes that of linear algebras 
%and by taking an inverse matrix, quasideterminants naturally appear.
%This is just a rewriting, however, quasideterminants
%have various identities such as quasi-Pl\"ucker relations

Furthermore investigation of the \NC{} extension of a bilinear form 
approach to the ASDYM equation
(Gilson, Nimmo and Ohta (1998);
Sasa, Ohta and Matsukidaira (1998);
 Wang and Wadati (2004))
would be beneficial
because many aspects in these paper are close to ours. 
%This might lead to a profound connection 
%between higher-dimensional integrable
%systems related to twistor theory and lower-dimensional ones
%related to Sato's theory.
%Extention to a higher-rank gauge group is worthwhile, for example,
%Yang equation of $U(3)$...
%We have found that quasideterminants play important roles in 
%4-dimensional the \NC{} ASDYM equation also, which
The relationship with \NC{} Darboux and 
\NC{} binary Darboux transformations (Salaam, Hassan and Siddiq (2007)) 
is also interesting.

\subsection*{Acknowledgments}

MH would like to thank L.~Mason and JJCN
for hospitality (and many helpful comments from LM) 
during stay at Mathematical Institute,
University of Oxford and at Department of Mathematics,
University of Glasgow, respectively.
The work of MH was supported by
%the Yamada Science Foundation
%for the promotion of the natural science,
Grant-in-Aid for Young Scientists (\#18740142),
the Nishina Memorial Foundation and the Daiko Foundation.

\begin{appendix}

\section{Brief review of quasideterminants}

In this section, we give a brief introduction to quasideterminants,
introduced by Gelfand and Retakh (2001),
in which a few of the key properties which play important roles
in Section 3 are described. 
More detailed discussion is seen in the survey (Gelfand, Gelfand,
Retakh and Wilson (2005)).

Quasideterminants are defined in terms of inverse matrices
and we suppose the existence of all matrix inverses referred to.
Let $A=(a_{ij})$ be an $n\times n$ matrix and  $B=(b_{ij})$ be
the inverse matrix of $A$, that is, $A B=B A =1$.
Here the matrix entries belong to a noncommutative ring.
Quasideterminants of $A$ are defined formally
as the inverses of the entries in $B$:
\begin{eqnarray}
\label{def_q}
\vert A \vert_{ij}:=b_{ji}^{-1}.
\end{eqnarray}
In the case that variables commute, this is reduced to
\begin{eqnarray}
 \vert A \vert_{ij}=
  (-1)^{i+j}\frac{\det A}{\det A^{ij}},
\label{limit}
\end{eqnarray}
where $A^{ij}$ is the matrix obtained from $A$ by
deleting the $i$-th row and the $j$-th column.

We can also write down a more explicit definition of quasideterminants.
In order to see this, let us recall the following formula
for the inverse a square $2\times 2$ block square matrix:
\begin{eqnarray*}
 \left[
 \begin{array}{cc}
  A&B \\C&d
 \end{array}
 \right]^{-1}
=\left[\begin{array}{cc}
A^{-1}+A^{-1} B S^{-1} C A^{-1}
 &-A^{-1} B S^{-1}\\
 -S^{-1} C A^{-1}
&S^{-1}
\end{array}\right],
\end{eqnarray*}
where $A$ is a square matrix, $d$ is a single element and $B$ and $C$ are column and row vectors of appropriate length and $S=d-C A^{-1} B$ is called a {\it Schur complement}. In fact this formula is valid for $A$, $B$, $C$ and $d$ in any ring not just for matrices. Thus the quasideterminant associated with the bottom right element is simply $S$. By choosing an appropriate partitioning, any entry in the inverse of a square matrix can be expressed as the inverse of a Schur complement and hence quasideterminants can also be defined recursively by:
\begin{eqnarray}
 \vert A \vert_{ij}=a_{ij}-\sum_{i^\prime (\neq i), j^\prime (\neq j)}
  a_{ii^\prime}  ((A^{ij})^{-1})_{i^\prime j^\prime} 
  a_{j^\prime
  j}
 =a_{ij}-\sum_{i^\prime (\neq i), j^\prime (\neq j)}
  a_{ii^\prime}  (\vert A^{ij}\vert_{j^\prime i^\prime })^{-1}
   a_{j^\prime j}.
\end{eqnarray}
It is sometimes convenient to use the following alternative notation in which a box is drawn about the corresponding entry in the matrix:
\begin{eqnarray}
 \vert A\vert_{ij}=
  \begin{array}{|ccccc|}
   a_{11}&\cdots &a_{1j} & \cdots& a_{1n}\\
   \vdots & & \vdots & & \vdots\\
   a_{i1}&~ & {\fbox{$a_{ij}$}}& ~& a_{in}\\
   \vdots & & \vdots & & \vdots\\
   a_{n1}& \cdots & a_{nj}&\cdots & a_{nn}
  \end{array}~.
\end{eqnarray}

Quasideterminants have various interesting properties
similar to those of determinants. Among them,
the following ones play important roles in this paper. In the block matrices given in these results, lower case letters denote single entries and upper case letters denote matrices of compatible dimensions so that the overall matrix is square. 

\begin{itemize}

 \item \NC{} Jacobi identity 

A simple and useful special case of the \NC{} Sylvester's Theorem
       (Gelfand and Retakh (1991)) is
\begin{equation}\label{nc syl}
    \begin{vmatrix}
      A&B&C\\
      D&f&g\\
      E&h&\fbox{$i$}
    \end{vmatrix}=
    \begin{vmatrix}
      A&C\\
      E&\fbox{$i$}
    \end{vmatrix}-
    \begin{vmatrix}
      A&B\\
      E&\fbox{$h$}
    \end{vmatrix}
    \begin{vmatrix}
      A&B\\
      D&\fbox{$f$}
    \end{vmatrix}^{-1}
    \begin{vmatrix}
      A&C\\
      D&\fbox{$g$}
    \end{vmatrix}.
\end{equation}

\item Homological relations (Gelfand and Retakh (1991))

\begin{eqnarray}\label{row hom}
    \begin{vmatrix}
      A&B&C\\
      D&f&g\\
      E&\fbox{$h$}&i
    \end{vmatrix}
&=&  \begin{vmatrix}
      A&B&C\\
      D&f&g\\
      E&h&\fbox{$i$}
    \end{vmatrix}
    \begin{vmatrix}
      A&B&C\\
      D&f&g\\
      0&\fbox{0}&1
    \end{vmatrix},\nonumber\\
\label{col hom}
    \begin{vmatrix}
      A&B&C\\
      D&f&\fbox{$g$}\\
      E&h&i
    \end{vmatrix}
&=&      \begin{vmatrix}
      A&B&0\\
      D&f&\fbox{0}\\
      E&h&1
    \end{vmatrix}
    \begin{vmatrix}
      A&B&C\\
      D&f&g\\
      E&h&\fbox{$i$}
    \end{vmatrix}
\end{eqnarray}

\item A derivative formula for quasideterminants (Gilson and Nimmo (2009))

\begin{align}
    \begin{vmatrix}
    A&B\\
    C&\fbox{$d$}
    \end{vmatrix}'
 \label{col diff} &=
    \begin{vmatrix}
    A&B'\\
    C&\fbox{$d'$}
    \end{vmatrix}
    +\sum_{k=1}^n
    \begin{vmatrix}
    A&(A_k)'\\
    C&\fbox{$(C_k)'$}
    \end{vmatrix}
    \begin{vmatrix}
    A&B\\
    e_k^t&\fbox{$0$}
    \end{vmatrix}
,
\end{align}
where $A_k$ is the $k$th column of a matrix $A$ and  
$e_k$ is the column $n$-vector $(\delta_{ik})$ (i.e.\ 1 in the
$k$th row and 0 elsewhere). 

\item A special formula of inverse of a quasideterminant (Gilson,
       Hamanaka and Nimmo (2007))

\begin{eqnarray}
\label{inverse}
\begin{vmatrix}
a&B&c&\alpha\\
D&E&F&0\\
g&H&\fbox{\hphantom0\llap{$i$}}&\fbox{0}\\
\beta&0&\fbox{0}&\fbox{0}\\
\end{vmatrix}^{-1}
=
\begin{vmatrix}
\fbox{0}&\fbox{0}&0&\gamma\\
\fbox{0}&\fbox{\hphantom0\llap{$a$}}&B&c\\
0&D&E&F\\
\delta&g&H&i
\end{vmatrix},
\end{eqnarray}
with $\alpha\beta=\gamma\delta=-1, \alpha+\gamma=0$,
where lower case letters denote single entries, upper case letters
denote matrices of compatible dimensions and Greek letters are
scalars (i.e. commute with everything). 

\end{itemize}

\end{appendix}


\baselineskip 5mm

\begin{thebibliography}{99}
%\small

\bibitem{AtWa}
  Atiyah, M.~F.~and Ward, R.~S. 1977
  Instantons and algebraic geometry.
  {\it Commun.  Math.  Phys.   } {\bf 55}, 117.
  %%CITATION = CMPHA,55,117;%%

\bibitem{Brain}
Brain, S.~J. 2005
The noncommutative Penrose-Ward transform
and self-dual Yang-Mills fields.
{\it Ph.  D Thesis}. University of Oxford.

\bibitem{BrMa}
  Brain, S.~J.~and Majid, S.~2008
  Quantisation of twistor theory by cocycle twist.
  {\it Commun.  Math.  Phys. }  {\bf 284}, 713.
  %%CITATION = MATH/0701893;%%

\bibitem{CLMS}
 Correa, D.~H., Lozano, G.~S., Moreno, E.~F. and Schaposnik, F.~A. 2001
  Comments on the $U(2)$ noncommutative instanton.
  {\it Phys.  Lett.   B} {\bf 515}, 206.
  %%CITATION = PHLTA,B515,206;%%

\bibitem{CoFa}
Corrigan, E. and Fairlie, D.~B. 1977
Scalar field theory and exact solutions to 
a classical $SU(2)$ gauge theory.
{\it Phys.  Lett.  B} {\bf 67}, 69.
%%CITATION = PHLTA,B67,69;%%

\bibitem{CFGY}
  Corrigan, E., Fairlie, D.~B., Yates, R.~G. and Goddard, P. 1978
  Backlund transformations and the construction
  of the Atiyah-Ward ansatz for selfdual $SU(2)$ gauge fields.
  {\it Phys.  Lett.  B} {\bf 72}, 354;
  %%CITATION = PHLTA,B72,354;%%
  The construction of selfdual solutions to $SU(2)$ gauge theory.
  {\it Commun.  Math.  Phys. }  {\bf 58}, 223.
  %%CITATION = CMPHA,58,223;%%

\bibitem{deVega}
  de Vega, H.~J. 1988
  Nonlinear multiplane wave solutions of selfdual Yang-Mills theory.
  {\it Commun.  Math.  Phys. }  {\bf 116}, 659.
  %%CITATION = CMPHA,116,659;%%

\bibitem{DiMH_proc}
 Dimakis, A. and M\"uller-Hoissen, F. 2004
 Extension of Moyal-deformed hierarchies of soliton equations.
  nlin/0408023.
  %%CITATION = NLIN/0408023;%%

\bibitem{DiMH}
  Dimakis, A. and M\"uller-Hoissen, F. 2007
  With a Cole-Hopf transformation to solutions of the noncommutative
  KP hierarchy in terms of Wronski matrices.
  {\it J.  Phys.  A}  {\bf 40}, F321.
%%CITATION = NLIN-SI 0701052;%%

\bibitem{DoNe}
 Douglas, M.~R. and Nekrasov, N.~A. 2002
  Noncommutative field theory.
 {\it Rev.  Mod.  Phys. }  {\bf 73}, 977.
%%CITATION = HEP-TH 0106048;%%

\bibitem{EGR}
 Etingof, P.,  Gelfand, I.~ and Retakh, V. 1997
  Factorization of differential operators,
  quasideterminants, and nonabelian Toda field equations.
  {\it Math.  Res.  Lett. } {\bf 4}, 413.
  %[q-alg/9701008].
  %%CITATION = Q-ALG 9701008;%%

\bibitem{EGR2}
  Etingof, P., Gelfand, I. and Retakh, V. 1998
  Nonabelian integrable systems, quasideterminants, and Marchenko lemma.
  {\it Math.  Res.  Lett. } {\bf 5}, 1.
  %[q-alg/9707017].
  %%CITATION = Q-ALG 9707017;%%

\bibitem{GeRe}
 Gelfand, I. and Retakh, V. 1991
 Determinants of matrices over noncommutative rings.
 {\it Funct.  Anal.  Appl. } {\bf 25}, 91.

\bibitem{GeRe2}
 Gelfand, I. and Retakh, V. 1992
 Theory of noncommutative determinants,
 and characteristic functions of graphs.
 {\it Funct.  Anal.  Appl. } {\bf 26}, 231.

\bibitem{GGRW}
  Gelfand, I., Gelfand, S., Retakh, V. and Wilson, R.~L. 2005
  Quasideterminants.
  {\it Adv.  Math. } {\bf 193}, 56.
  %[math.qa/0208146].
  %%CITATION = MATH-QA 0208146;%%
 
\bibitem{GiNi}
 Gilson, C.~R., and Nimmo, J.~J.~C. 2007
 On a direct approach to quasideterminant solutions of
 a noncommutative KP equation.
{\it J.  Phys.  A} {\bf 40}, 3839.
%[nlin.si/0701027].
%%CITATION = NLIN-SI 0701027;%%

\bibitem{GHN}
  Gilson, C.~R., Hamanaka, M. and Nimmo, J.~J.~C. 2009
  B\"acklund transformations for noncommutative anti-self-dual
  Yang-Mills equations.
  {\it Glasgow Mathematical Journal} {\bf 51A}, 83.
  %[arXiv:0709.2069].
  %%CITATION = ARXIV:0709.2069;%%

\bibitem{GiMa}
 Gilson, C.~R., and Macfarlane, S.~R. 2009
 Dromion solutions of noncommutative Davey-Stewartson equations
{\it J. Phys. A} {\bf 42}, 235202. 
%[arXiv:0901.4918].
%%CITATION = ARXIV:0901.4918;%%

\bibitem{GNO98}
   Gilson, C.~R., Nimmo, J.~J.~C. and Ohta, Y. 1998
  ``Self dual Yang Mills and bilinear equations. 
in {\it Recent Developments in Soliton Theory}, edited by M. Oikawa,
Kyushu University, 147-153.

\bibitem{GNO}
 Gilson, C.~R., Nimmo, J.~J.~C. and Ohta, Y. 2007
Quasideterminant solutions of a non-Abelian Hirota-Miwa equation.
{\it J.\ Phys.\ A} {\bf 40}, 12607.
%[nlin.SI/0702020].
%%CITATION = NLIN-SI 0702020;%%

\bibitem{GNS}
 Gilson, C.~R., Nimmo, J.~J.~C. and Sooman, C.~M. 2008a
   On a direct approach to quasideterminant solutions of a
   noncommutative modified KP equation. 
  {\it J.\ Phys.\ A} {\bf 41}, 085202.
  %[arXiv:0711.3733].
  %%CITATION = ARXIV:0711.3733;%%

\bibitem{GNS2}
 Gilson, C.~R., Nimmo, J.~J.~C. and Sooman, C.~M. 2008b
  Matrix solutions of a noncommutative KP equation and 
   a noncommutative mKP equation. 
  arXiv:0810.1891.
  %%CITATION = ARXIV:0810.1891;%%

\bibitem{GoVe}
 Goncharenko, V.~M. and Veselov, A.~P. 1998
 Monodromy of the matrix Schrodinger equations and Darboux
 transformations.
 {\it J.  Phys.  A} {\bf 31}, 5315.
 %%CITATION = JPAGB,A31,5315;%%

%\bibitem{GiGu}
%Gu, F.  
%{\it private communication}.
% ``,''

\bibitem{HaHa}
  Haider, B. and Hassan, M. 2008
  The $U(N)$ chiral mode and exact multi-solitons.
  {\it J.  Phys.  A}  {\bf 41}, 255202.
  %%CITATION = JPAGB,A41,255202;%%
	
\bibitem{Hamanaka_PhD}
Hamanaka, M. 2003
Noncommutative solitons and D-branes.
{\it Ph. D thesis}. % (University of Tokyo, 2003)
%hep-th/0303256.
%%CITATION = HEP-TH 0303256;%%

\bibitem{Hamanaka_proc}
Hamanaka, M. 2005a
Noncommutative solitons and integrable systems.
%in {\it Noncommutative Geometry and Physics}, edited by
%Y.~Maeda, N.~Tose, N.~Miyazaki, S.~Watamura and D.~Sternheimer
%(World Sci., 2005) 175
hep-th/0504001.
%%CITATION = HEP-TH 0504001;%%

\bibitem{Hamanaka05_PLB}
 Hamanaka, M., 2005b
 On reductions of noncommutative anti-self-dual Yang-Mills
  equations.
 {\it Phys.  Lett.  B} {\bf 625}, 324.
 %[hep-th/0507112].
 %%CITATION = HEP-TH 0507112;%%

\bibitem{Hamanaka06_NPB}
  Hamanaka, M. 2006
  Noncommutative Ward's conjecture and integrable systems.
  {\it Nucl.  Phys.  B} {\bf 741}, 368.
  %[hep-th/0601209].
  %%CITATION = HEP-TH 0601209;%%

 \bibitem{Hamanaka06_JHEP}
  Hamanaka, M. 2007
  Notes on exact multi-soliton solutions of
   noncommutative integrable hierarchies.
  {\it JHEP} {\bf 0702}, 094.
  %[hep-th/0610006].
  %%CITATION = HEP-TH 0610006;%%

\bibitem{HaTo}
  Hamanaka, M.  and Toda, K. 2003
  Towards noncommutative integrable equations.
%{\it In the Proceedings of 5th International Conference on Symmetry 
%in Nonlinear Mathematical Physics (SYMMETRY 03), Kiev, Ukraine, 
%23-29 Jun 2003, pp 404-411}
  {[hep-th/0309265]}.
  %%CITATION = ECONF,C0306234,404;%%

\bibitem{Hannabuss}
Hannabuss, K.~C. 2001
Non-commutative twistor space.
{\it Lett.  Math.  Phys. }  {\bf 58}, 153.
%[hep-th/0108228].
%%CITATION = HEP-TH 0108228;%%

\bibitem{Hassan}
  Hassan, M. 2009
  Darboux transformation of the generalized coupled dispersionless
  integrable system. 
{\it J.  Phys.  A} {\bf 42}, 065203.

%\bibitem{Hirota}
%Hirota, R. (translated by Gilson, C.~R., Nagai, A. and Nimmo, J.~J.~C.) 2004
%{\it The Direct Methods in Soliton Theory},
%Cambridge UP%[ISBN/0521836603].

\bibitem{HLW}
Horv\'ath, Z., Lechtenfeld, O. and Wolf, M. 2002
Noncommutative instantons via dressing and splitting approaches.
{\it JHEP} {\bf 0212}, 060.
%[hep-th/0211041].
%%CITATION = HEP-TH 0211041;%%

\bibitem{IhUh}
  Ihl, M. and Uhlmann, S. 2003
  Noncommutative extended waves and soliton-like
  configurations in N = 2 string theory.
  {\it Int.  J.  Mod.  Phys.   A} {\bf 18}, 4889.
  %[hep-th/0211263].
  %%CITATION = IMPAE,A18,4889;%%

\bibitem{KKO}
Kapustin, A., Kuznetsov, A. and Orlov, D. 2001
Noncommutative instantons and twistor transform.
{\it Commun.  Math.  Phys. }  {\bf 221}, 385.
%[hep-th/0002193].
%%CITATION = HEP-TH 0002193;%%

\bibitem{Kupershmidt}
Kupershmidt, B. 2000
{\it KP or mKP},
Mathematical Surveys and Monographs vol 78. AMS.
%[ISBN/0821814001].

\bibitem{Lechtenfeld_proc}
Lechtenfeld, O. 2008
%Noncommutative solitons.
%hep-th/0605034;
%%CITATION = HEP-TH 0605034;%%
Noncommutative solitons.
  AIP Conf.\ Proc.\  {\bf 977}, 37.
  %[arXiv:0710.2074].
  %%CITATION = APCPC,977,37;%%

\bibitem{LePo02}
Lechtenfeld, O. and Popov, A.~D. 2002
Noncommutative 't Hooft instantons.
{\it JHEP} {\bf 0203}, 040.
%[hep-th/0109209].
%%CITATION = HEP-TH 0109209;%%

\bibitem{LPS}
  Lechtenfeld, O., Popov, A.~D. and Spendig, B. 2001
  Open N = 2 strings in a B-field background and noncommutative
  self-dual Yang-Mills.
  {\it Phys.  Lett. }  B {\bf 507}, 317.
  %[hep-th/0012200].
  %%CITATION = PHLTA,B507,317;%%

\bibitem{LPS2}
  Lechtenfeld, O., Popov, A.~D. and Spendig, B. 2001
  Noncommutative solitons in open N = 2 string theory.
  {\it JHEP} {\bf 0106}, 011.
  %[hep-th/0103196].
  %%CITATION = JHEPA,0106,011;%%

\bibitem{LiNi}
  Li, C.~X. and Nimmo, J.~J.~C. 2008
  Quasideterminant solutions of a non-Abelian Toda lattice 
  %  and kink solutions of a matrix sine-Gordon equation.
  {\it Proc.  Roy.  Soc.  Lond.   A} {\bf 464}, 951.
  %[arXiv:0711.2594].
  %%CITATION = ARXIV:0711.2594;%%

\bibitem{LiNi2}
  Li, C.~X. and Nimmo, J.~J.~C. 2009
  A noncommutative semi-discrete Toda equation 
   and its quasideterminant solutions.
  {\it Glasgow Math. J.} {\bf 51A} 121.
  %%CITATION = ARXIV:0806.3598;%%

\bibitem{LNT}
  Li, C.~X., Nimmo, J.~J.~C. and Tamizhmani, K.~M. 2009
  On solutions to the non-Abelian Hirota-Miwa equation 
    and its continuum limits.
  {\it Proc. R. Soc. A} {\bf 465}, 1441.
  %%CITATION = ARXIV:0809.3833;%%

 \bibitem{CMN}
  Mason, L., Chakravarty, S. and ~Newman, E.~T. 1988
  Backlund transformations for the antiselfdual Yang-Mills equations.
  {\it J.  Math.  Phys. }  {\bf 29}, 1005;
  %%CITATION = JMAPA,29,1005;%%
  A simple solution generation method for antiselfdual Yang-Mills
  equations.
  {\it Phys.  Lett.  A} {\bf 130}, 65.
  %%CITATION = PHLTA,A130,65;%%

\bibitem{MaWo}
Mason, L.~J. and Woodhouse, N.~M. 1996
{\it Integrability, Self-Duality, and Twistor Theory},
Oxford UP.
%(London Mathematical Society monographs, new series: 15)},
%[ISBN/0-19-853498-1].

\bibitem{Mazzanti}
  Mazzanti, L. 2007
  Topics in noncommutative integrable theories and holographic brane-world
  cosmology.
  {\it Ph.  D thesis}.
  %%CITATION = ARXIV:0712.1116;%%

\bibitem{NeSc}
  Nekrasov, N.  and Schwarz, A. 1998
  Instantons on noncommutative $R^4$ and
  (2,0) superconformal six  dimensional theory.
  {\it Commun.  Math.  Phys.  } {\bf 198}, 689.
  %[hep-th/9802068].
  %%CITATION = HEP-TH 9802068;%%

\bibitem{Nimmo}
 Nimmo, J.~J.~C. 2006
 On a non-Aberian Hirota-Miwa equation.
 {\it J.  Phys.  A} {\bf 39}, 5053.
 %%CITATION = JPAGB,A39,5053;%%

\bibitem{GNO00}
  Nimmo, J.~J.~C. , Gilson, C.~R. and Ohta, Y. 2000
  Applications of Darboux transformations
  to the selfdual Yang-Mills equations.
  {\it Theor.  Math.  Phys. }  {\bf 122}, 239.
  %[Teor.  Mat.  Fiz.   {\bf 122}, 284 (2000)].
  %%CITATION = TMPHA,122,239;%%

\bibitem{HSS07}
  Saleem, U., Hassan, M. and Siddiq, M. 2007
  Non-local continuity equations and binary Darboux transformation
        of noncommutative (anti) self-dual Yang-Mills equations.
  {\it J.  Phys.  A}  {\bf 40}, 5205.
  %%CITATION = JPAGB,A40,5205;%%

\bibitem{SaPe}
 Samsonov, B.~F. and Pecheritsin, A.~A. 2004
 Chains of Darboux transformations for the matrix Schr\"odinger equation.
 {\it J.  Phys.  A} {\bf 37}, 239.
 %%CITATION = JPAGB,A37,239;%%

\bibitem{MOS}
  Sasa, N., Ohta, Y. and Matsukidaira, J. 1998
  Bilinear form approach to the self-dual Yang-Mills equation
  and integrable system in (2+1)-dimension.
  {\it J.  Phys.  Soc.  Jap. }  {\bf 67}, 83.
  %%CITATION = JUPSA,67,83;%%

\bibitem{Szabo}
Szabo, R.~J. 2003
Quantum field theory on noncommutative spaces.
{\it Phys.  Rept. }  {\bf 378}, 207.
%[hep-th/0109162].
%%CITATION = HEP-TH 0109162;%%

\bibitem{Takasaki}
Takasaki, K. 2001
Anti-self-dual Yang-Mills equations on noncommutative spacetime.
{\it J.  Geom.  Phys. } {\bf 37}, 291.
%[hep-th/0005194].
%%CITATION = HEP-TH 0005194;%%

\bibitem{Tamassia}
Tamassia, L.
Noncommutative supersymmetric / integrable models 
and string theory.
{\it Ph.  D thesis}.% hep-th/0506064.
%%CITATION = HEP-TH 0506064;%%

\bibitem{tHooft}
't Hooft, G. 1976 {\it unpublished}.

\bibitem{WaWa}
  Wang, N. and Wadati, M. 2004
  Noncommutative KP hierarchy and Hirota triple-product relations.
  {\it J.  Phys.  Soc.  Jap. }  {\bf 73}, 1689.
  %%CITATION = JUPSA,73,1689;%%

\bibitem{Wilczek}
Wilczek, F. 1977
Geometry and interactions of instantons.
in %{\it C76-06-14.3.4}
%Print-76-0718 (PRI\NC{}ETON)
%{\it Loosely based on talk given
%at Univ. of Rochester Conf. on Quark Binding,
%Rochester, N.Y., Jun 14-18, 1976}.
{\it Quark Confinement and Field Theory}
edited by D. R. Stump and D. H. Weingarten 
(John Wiley \& Sons, USA), 211.
%[ISBN/0-471-02721-9]}.

\bibitem{Yang}
  Yang, C.~N. 1977
  Condition of selfduality for $SU(2)$ gauge fields on Euclidean
  four-dimensional space.
  {\it Phys.  Rev.  Lett. }  {\bf 38}, 1377.
  %%CITATION = PRLTA,38,1377;%%

 
\end{thebibliography}
\end{document}